\documentclass[11pt,letterpaper]{article}
\usepackage[top=1in,bottom=1in,left=1in,right=1in]{geometry}

\usepackage{microtype}
\usepackage{xcolor}
\usepackage{amsmath}
\usepackage{amssymb}
\usepackage{amsthm}
\usepackage{mathrsfs}
\usepackage{enumitem}
\usepackage{textgreek}
\usepackage{graphicx}
\usepackage{framed}
\usepackage{doi}
\usepackage[textsize=tiny]{todonotes}
\usepackage[framemethod=tikz]{mdframed}
\usepackage{thm-restate}
\usepackage{mathtools}
\usepackage{xspace}
\usepackage[capitalize, nameinlink]{cleveref}
\usepackage{multirow}
\usepackage{cellspace}
\usepackage{array}
\usepackage{tcolorbox}
\usepackage[sort&compress]{natbib}

\usepackage{hyperref}
\hypersetup{
    colorlinks = true,
    allcolors = black
}

\Crefname{remark}{Remark}{Remarks}
\Crefname{observation}{Observation}{Observations}

\theoremstyle{plain}
\newtheorem{theorem}{Theorem}[section]
\newtheorem{lemma}[theorem]{Lemma}

\newtheorem{corollary}[theorem]{Corollary}

\newtheorem{observation}[theorem]{Observation}

\theoremstyle{definition}
\newtheorem{definition}[theorem]{Definition}

\theoremstyle{plain}

\newcounter{open}
\newtheorem{oq}[open]{Open Problem}

\theoremstyle{remark}

\newcommand{\namedref}[2]{\hyperref[#2]{#1~\ref*{#2}}}

\newcommand{\fA}{\mathcal A}

\newcommand{\fC}{\mathcal C}
\newcommand{\fE}{\mathcal E}
\newcommand{\fI}{\mathcal I}
\newcommand{\fJ}{\mathcal J}
\newcommand{\fK}{\mathcal K}
\newcommand{\fL}{\mathcal L}

\newcommand{\fR}{\mathcal R}
\newcommand{\fS}{\mathcal S}
\newcommand{\fT}{\mathcal T}
\newcommand{\fW}{\mathcal W}
\newcommand{\fX}{\mathcal X}

\newcommand{\Gz}{G^{(0)}}
\newcommand{\Go}{G^{(1)}}

\newcommand{\Gi}{G^{(i)}}

\newcommand{\vo}{v^{(1)}}
\newcommand{\vt}{v^{(2)}}
\newcommand{\vi}{v^{(i)}}
\newcommand{\vj}{v^{(j)}}

\DeclareMathOperator{\len}{\operatorname{len}}

\newcommand{\minusspace}{\vspace{-0.3cm}}

\definecolor{darkgreen}{rgb}{0,0.5,0}
\definecolor{darkred}{rgb}{0.4,0,0}

\title{A Post-Quantum Lower Bound for the Distributed Lov\'asz Local Lemma}
\author{Sebastian Brandt\footnote{CISPA Helmholtz Center for Information Security}\and Tim Göttlicher\footnote{Saarland University, CISPA Helmholtz Center for Information Security}}
\date{}

\begin{document}
\maketitle

 \setcounter{page}{0}
 \thispagestyle{empty}

\begin{abstract}
    In this work, we study the Lov\'asz local lemma (LLL) problem in the area of \emph{distributed} quantum computing, which has been the focus of attention of recent advances in quantum computing [STOC'24, STOC'25, STOC'25].
    We prove a lower bound of $2^{\Omega(\log^* n)}$ for the complexity of the distributed LLL in the quantum-LOCAL model.
    More specifically, we obtain our lower bound already for a very well-studied special case of the LLL, called sinkless orientation, in a stronger model than quantum-LOCAL, called the randomized online-LOCAL model.
    As a consequence, we obtain the same lower bounds for sinkless orientation and the distributed LLL also in a variety of other models studied across different research communities.

    Our work provides the first superconstant lower bound for sinkless orientation and the distributed LLL in all of these models, addressing recently stated open questions.
    Moreover, to obtain our results, we develop an entirely new lower bound technique that we believe has the potential to become the first generic technique for proving post-quantum lower bounds for many of the most important problems studied in the context of locality.
\end{abstract}

\clearpage

\section{Introduction}\label{sec:intro}
In this work, we study the task of proving lower bounds for quantum algorithms in the distributed setting, with a focus on the Lov\'asz local lemma.
In recent years, distributed quantum computing has seen a surge of exciting results making considerable progress in a variety of settings, such as in the quantum versions of the CONGEST and Congested Clique models which capture bandwidth-constrained distributed algorithms~\cite{le2018sublinear,izumi2019quantum,IzumiGM20,wang2021complexity,magniez2022quantum,CHFG22,van2022framework,wu2022quantum,fraigniaud2024even,dufoulon2025quantum} and other models~\cite{gavoille2009can,arfaoui2014can,elkin2014can,LGNR19quantum,gall2022non,coiteux2024no,hasegawa2024power,dhar2024local,akbari2025online,balliu2025distributed}.
Some of the most recent of these works study the quantum version of the standard LOCAL model of distributed computation (called quantum-LOCAL), achieving a number of breakthrough results for locally checkable problems\footnote{Locally checkable problems are problems that can be defined via local (i.e., constant-hop) constraints, i.e., problems for which a claimed solution is correct if and only if the constant-hop neighborhood of each node satisfies some problem-specific local criteria. This problem class contains most of the problems studied in the LOCAL model, such as coloring problems, maximal independent set, maximal matching, and the Lov\'asz local lemma problem.}.
For instance, \cite{coiteux2024no} shows that for approximate graph coloring problems, there is essentially no quantum advantage in the LOCAL model, by proving upper bounds that hold in LOCAL and lower bounds that hold in quantum-LOCAL and match the upper bounds up to a polylogarithmic factor.
In contrast, \cite{balliu2025distributed} shows that there exist locally checkable problems for which quantum computation helps in the LOCAL model, by proving a separation between the LOCAL and quantum-LOCAL complexities for a locally checkable problem called ``iterated GHZ''.

Besides the study of quantum advantage, another topic of considerable interest is the relation of quantum-LOCAL to other models related to locality.
In~\cite{akbari2025online}, the authors show that there are fundamental connections between a variety of such models that have been studied across different fields (and include quantum-LOCAL).
Before delving into the details of this work, we introduce the two main models discussed above.
For the reader unfamiliar with the quantum world, we already note at this point that no deep understanding of quantum computation is necessary for obtaining our lower bound, since most technical details are examined in a setting that can be studied independently of any quantum considerations.

\minusspace
\paragraph{The LOCAL model.}
In the \emph{LOCAL model} of distributed computation~\cite{linial1992locality,peleg2000}, each vertex (or node) of the input graph $G$ for some given graph problem is considered as a computational entity.
These entities can exchange messages via incident edges; more precisely, computation/communication proceeds in synchronous rounds, and in each round each node first sends an arbitrarily large message to each of its neighbors, then receives the messages sent by its neighbors, and finally can perform some (unbounded) internal computation on the information it has accumulated so far.
In the beginning of the overall computation, each node $v$ does not know $G$; node $v$ merely knows its own degree $\deg(v)$, can distinguish between its incident edges via port numbers $1, \dots, \deg(v)$ uniquely assigned to these edges by $v$, and knows the total number $n$ of nodes of the input graph and some symmetry-breaking information that depends on whether the considered model is the \emph{deterministic} or \emph{randomized} LOCAL model.
In the deterministic LOCAL model, each node is equipped with an ID (in the range from $1$ to $n^c$ for some constant $c$) that is unique across the entire graph; in the \emph{randomized} LOCAL model, each node is assigned a private random bit string of unbounded length.
Each node executes the same algorithm that specifies which messages to send and which computations to perform.
Each node $v$ also has to decide to terminate at some point, upon which it provides its local part of the solution to the given problem.
The (time or round) complexity of an algorithm in the LOCAL model is simply the (worst-case) number of rounds it takes until the last node terminates; the complexity of a problem is the time complexity of an optimal algorithm for that problem.
In the randomized LOCAL model, an algorithm is required to produce a correct output with high probability (w.h.p.), i.e., with probability at least $1 - 1/n$.

\minusspace
\paragraph{The quantum-LOCAL model.}
The \emph{quantum-LOCAL model} (see, e.g., \cite{gavoille2009can,akbari2025online}) is essentially the randomized LOCAL model enhanced with quantum communication/computation.
More specifically, each node is a quantum computer holding an arbitrarily large number of quantum bits (qubits), which form local states that can be manipulated in the standard quantum manner via unitary transformations. 
Moreover, instead of exchanging standard messages consisting of bits, in quantum-LOCAL the messages sent and received by the nodes consist of qubits.
Finally, the nodes may perform arbitrary quantum measurements.
This concludes the description of the quantum-LOCAL model.
In a common variant of quantum-LOCAL, called ``quantum-LOCAL (shared)'', before the communication starts and the nodes receive their initial information, the algorithm may additionally provide the nodes with shared information, including a shared entangled quantum state.
We note that the lower bounds that we prove apply to both versions of the model.
However, for simplicity, we will mainly consider quantum-LOCAL throughout the paper and avoid mentioning repeatedly that our results also hold in quantum-LOCAL (shared). 

\minusspace
\paragraph{A landscape of locality models.}
Following the results from~\cite{akbari2025online} (and earlier works studying relations between models, such as~\cite{gavoille2009can,akbari2023locality}), a highly interesting picture capturing relations between all kinds of models related to locality has emerged.
This landscape includes standard distributed models such as LOCAL or quantum-LOCAL, models based on probability distributions such as non-signaling or bounded-dependence distributions which have been studied extensively across different communities (see, e.g., \cite{gavoille2009can,arfaoui2014can,holroyd2016finitely,holroyd2017finitary,holroyd2018finitely,holroyd2024symmetrization}), distributed sequential models such as the SLOCAL model~\cite{ghaffari2018derandomizing}, and centralized sequential models such as the dynamic-LOCAL and online-LOCAL models~\cite{akbari2023locality}.
The aforementioned relations between the models are usually statements of the form ``model A is at least as strong as model B'' which is to be understood in the sense that an algorithm in model B can be simulated in model A without (asymptotic) loss in runtime and failure probability.
An example of a sequence of models with increasing strength is randomized LOCAL, quantum-LOCAL, quantum-LOCAL (shared), non-signaling, randomized online-LOCAL.
In fact, the randomized online-LOCAL model can simulate all models considered in the landscape in the above sense, making it the strongest of those models.
For the full overview of all models and their relations, we refer the reader to~\cite{akbari2025online}, in particular to Fig.\ 10 therein.

\minusspace
\paragraph{Post-quantum distributed lower bounds and beyond.}
Keeping the above model landscape in mind, let us turn our attention back to the task of proving post-quantum lower bounds, i.e., LOCAL model lower bounds for locally checkable problems that do not break if we go to quantum-LOCAL.
As already indicated in~\cite{akbari2025online}, proving lower bounds directly in quantum-LOCAL seems to be beyond the current state-of-the-art techniques.
Instead, the known lower bounds that hold in quantum-LOCAL are achieved by proving the respective bounds in stronger models, in particular the non-signaling model (which, due to the aforementioned model relations then directly imply the same lower bound in quantum-LOCAL).

In particular, these bounds include bounds obtained via LOCAL model techniques that also work in the non-signaling model:
Via existential graph-theoretic arguments, \cite{coiteux2024no} showed a lower bound of $\Omega(n^{1/\alpha})$ for $c$-coloring $\chi$-chromatic graphs (where $\alpha := \lfloor\frac{c-1}{\chi-1}\rfloor$) in the non-signaling model and also extended Linial's logarithmic lower bound for coloring trees~\cite{linial1992locality} to this model.
Moreover, as observed in~\cite{gavoille2009can}, lower bounds in the (randomized) LOCAL model that are achieved via indistinguishability arguments such as an $\Omega(\sqrt{\frac{\log n}{\log \log n}})$-round lower bound for maximal matching and maximal independent set~\cite{kuhn2016local} and a $\Omega(\frac{\log n}{\log \log n})$-round lower bound for greedy coloring~\cite{gavoille2009complexity} transfer to the non-signaling model (called $\varphi$-LOCAL in~\cite{gavoille2009can}) and therefore also to the quantum-LOCAL model.
For locally checkable problems on constant-degree \emph{rooted trees}, \cite{akbari2025online} proved a general lifting theorem showing that any $\omega(\log^* n)$-round lower bound in the (deterministic or randomized) LOCAL model implies an $\Omega(\log \log \log n)$ lower bound (for the same problem) in the randomized online-LOCAL model (and therefore also in quantum-LOCAL).
Furthermore, \cite{akbari2025online} showed an $\Omega(\log n)$-round lower bound for $3$-coloring $2$-dimensional grids in the randomized online-LOCAL model (building on the same lower bound in the deterministic online-LOCAL model from~\cite{chang2024tight}); note however, that the implications for quantum-LOCAL are subsumed by the aforementioned results by~\cite{coiteux2024no}.

In general, the currently available techniques for proving lower bounds for locally checkable problems in quantum-LOCAL and stronger models are quite limited: there are lower bound tools such as indistinguishability arguments or existential graph theoretic arguments that are applicable to the quantum-LOCAL and non-signaling models but they do not provide anything close to a lower bound technique that is widely applicable and able to capture the most important problems such as coloring problems or the Lov\'asz local lemma problem.
This is in sharp contrast to the situation in the LOCAL model where such a technique is available in the form of the round elimination technique~\cite{brandt2016lower,brandt2019automatic}
and raises the following fundamental question.

\vspace{0.05cm}
\begin{tcolorbox}
    \begin{oq}\label{op1}
        How can we develop a \emph{generic} lower bound technique for locally checkable problems in the quantum-LOCAL model (and stronger models)?
    \end{oq}
\end{tcolorbox}
\vspace{0.05cm}

A natural first candidate for such a technique would be the aforementioned round elimination technique; however, unfortunately, \cite{balliu2025distributed} recently showed that round elimination cannot be used to prove lower bounds for quantum-LOCAL.
We also remark that, in terms of applicability to many models at once, a lower bound technique as desired in~\Cref{op1} would ideally already apply to the strongest of the models in the aforementioned model landscape---the randomized online-LOCAL model.

\minusspace
\paragraph{Fundamental problems.}
While obtaining a generic lower bound technique is a highly ambitious general goal, there are also many concrete problems for which close to nothing is known w.r.t.\ proving lower bounds in quantum-LOCAL and beyond.
Two of the most important such problems are the $(\Delta + 1)$-coloring problem and the Lov\'asz local lemma problem, which are both used as subroutines in countless algorithms for other problems in the LOCAL model (and stronger models).

The $(\Delta + 1)$-coloring problems asks to color the nodes of the input graph $G$ with $\Delta + 1$ colors such that no two adjacent nodes receive the same color, where $\Delta$ denotes the maximum degree of $G$.
In the LOCAL model, a lower bound of $\Omega(\log^* n)$ rounds\footnote{The expression $\log^* n$ denotes the smallest number of times the $\log$-function has to be applied iteratively to $n$ to obtain a value that is at most $1$.} is known for $(\Delta + 1)$-coloring which is tight for the special case of $\Delta = 2$, i.e., for the problem of $3$-coloring a cycle or path~\cite{cole1986deterministic,linial1992locality}.
However, in quantum-LOCAL even the latter special case is completely open: essentially nothing is known besides the fact that $3$-coloring on cycles/paths can be solved in $O(\log^* n)$ rounds (which trivially follows from the same upper bound in LOCAL).
Fascinatingly, resolving this coloring problem was one of the earliest results in the LOCAL model, proved more than three decades ago, which emphasizes our lack of understanding of lower bounds in quantum-LOCAL.

\minusspace
\paragraph{Lov\'asz local lemma.} 
The other of the two mentioned problems, the Lov\'asz local lemma problem, is a fundamental algorithmic problem that goes back to an existential probabilistic statement proved in 1975 by Erd{\H o}s and Lov\'asz~\cite{erdos75local}.
This statement, called the Lov\'asz local lemma (LLL), can be phrased as follows.

\begin{lemma}[LLL]\label{lem:existentiallll}
    Let $\fE = \{ E_1, \dots, E_k \}$ be a finite set of probabilistic events, $d$ a positive integer, and $p < 1$ a nonnegative real number such that, for each $1 \leq i \leq k$,
    \begin{enumerate}
        \item the event $E_i$ is mutually independent of all other $E_j$ except up to $d$ many, and
        \item the probability that $E_i$ occurs is upper bounded by $p$.
    \end{enumerate}
    Then, if $4pd \leq 1$, the probability that none of the events in $\fE$ occurs is positive.
\end{lemma}

The criterion $4pd \leq 1$ under which~\Cref{lem:existentiallll} holds was subsequently slightly relaxed by works of Spencer~\cite{Spencer77} and Shearer~\cite{shearer85spencer}.

The \emph{algorithmic LLL} (or \emph{constructive LLL}) is the algorithmic problem of finding a point in the probability space of the LLL setting that avoids all of the events in $\fE$.
Initiated by Beck~\cite{beck91algorithmic}, the algorithmic LLL has received considerable attention since then (see, e.g., \cite{alon91parallel,molloy98further,czumaj00algorithmic,moser08derandomizing,srinivasan08improved,moser09constructive,moser10constructive,haeupler10constructive,chandrasekaran13deterministic}), including a celebrated result by Moser and Tardos presenting an efficient randomized algorithm for solving the algorithmic LLL.

\minusspace
\paragraph{The distributed LLL.}
In the distributed setting, the algorithmic LLL is modeled as a graph problem as follows.
Let $\fX = \{ X_1, \dots, X_{\ell} \}$ be a finite set of mutually independent random variables, $\fE = \{ E_1, \dots, E_k \}$ a set of probabilistic events defined on $\fX$, and $p < 1$ a nonnegative real number such that, for each of the events in $\fE$, the probability that it occurs is upper bound by $p$.
Then, the dependency graph $G$ of this LLL instance (which constitutes the input graph to the distributed LLL), is defined by setting the node set to be $\fE$ and connecting two nodes $E_i, E_j$ by an edge if there is a common random variable from $\fX$ that both events depend on.
Let $d$ be an upper bound for the maximum degree of $G$, and assume that $p$ and $d$ satisfy some LLL criterion (e.g., $4pd \leq 1$) that guarantees the existence of a point in the probability space avoiding all events.

In the distributed LLL, the task is to find a solution to the algorithmic LLL in the respectively considered distributed setting (e.g., in LOCAL or quantum-LOCAL).
More precisely, each node with corresponding event $E_i$ has to output an assignment to all variables from $\fX$ that $E_i$ depends on such that
\begin{enumerate}
    \item $E_i$ does not occur under this assignment, and
    \item all nodes assigning a value to the same random variable assign the same value to the random variable.
\end{enumerate}

Following the aforementioned result by Moser and Tardos, which also yields an $O(\log^2 n)$-round algorithm in the LOCAL model, the distributed LLL was studied in an abundance of works~\cite{chung2014distributed,brandt2016lower,ghaffari2016improved,FischerG17,ghaffari2018derandomizing,chang2019exponential,chang2019time,chang2019distributed,brandt2019sharp,rozhovn2020polylogarithmic,brandt2020generalizing,MausU21,davies2023improved,halldorsson2024distributed,daviespeck2025on}, improving the state of the art in a variety of settings, for different LLL criteria.
For instance, in the randomized LOCAL model, the current state of the art is $O(\log d \cdot \log_{1/(ep(d + 1))} n)$ rounds~\cite{chung2014distributed} for the (most popular) LLL criterion $ep(d + 1) < 1$ and $O(\min\{ \log_{1/p} n, d/\log d + \log^{O(1)} \log n \})$ rounds~\cite{chung2014distributed,davies2023improved} for polynomial LLL criteria, i.e., criteria of the form $p\cdot d^c \in O(1)$ for some constant $c$ (which have been considered frequently in the distributed context).
To the best of our knowledge, no better upper bounds are known for any of the stronger models from~\cite[Fig.\ 10, arxiv version]{akbari2025online.stoc}.
On the lower bound side, the current state-of-the-art complexity bound in the randomized LOCAL model is $\Omega(\log \log n)$ rounds~\cite{brandt2016lower} (which holds for a wide variety of LLL criteria), which is achieved by showing this lower bound for a special case of the algorithmic LLL, called sinkless orientation.

\minusspace
\paragraph{Sinkless orientation.}
The sinkless orientation problem, introduced in~\cite{brandt2016lower} in the distributed setting, asks to orient the edges of the input graph such that no node of degree at least $3$ is a sink, i.e., such that each node of degree at least $3$ has at least one outgoing edge.
More specifically, the output for each node $v$ is an orientation of the edges incident to $v$ such that at least one of these edges is oriented away from $v$ and adjacent nodes agree on the orientation of the connecting edge.

When phrasing sinkless orientation in the language of LLL, each edge is considered to be a random variable (whose values are the two orientations of the considered edge) and the event associated with each node is that the node is a sink and has at least $3$ incoming edges.
Thereby, finding a solution to sinkless orientation is equivalent to solving this particular LLL problem.

If we allow that the input graph contains nodes of degree precisely $3$, then no LLL criterion guaranteeing the existence of a solution is satisfied (even though a solution to sinkless orientation always exists, also in this case).
However, as soon as we consider graphs with maximum degree $\Delta \geq 4$ where every node of degree at least $3$ has degree precisely $\Delta$,
sinkless orientation becomes a special case of LLL as setting $p := 1/2^{\Delta}$ and $d := \Delta$ satisfies, e.g., the LLL criterion $4pd \leq 1$.
In fact, sinkless orientation satisfies the very restrictive \emph{exponential} LLL criterion $p \cdot 2^d \leq 1$, which is especially interesting from a lower bound perspective, as any lower bound for LLL proved under this criterion also applies to all more permissive criteria, including all the commonly used criteria.
When discussing our results, we will implicitly assume this exponential LLL criterion without explicitly repeating this fact.
We also remark that the restriction to the aforementioned graph class makes the bounds proved in this work only stronger as our objective is to prove lower bounds.

In contrast to the situation for the distributed LLL, an upper bound of $O(\log \log n)$ rounds matching the lower bound is known in the randomized LOCAL model.
While no better upper bounds are known for the models sandwiched between the randomized LOCAL model and the randomized online-LOCAL model, in the randomized online-LOCAL model itself an upper bound of $O(\log \log \log n)$
is known, following from~\cite{ghaffari2017distributed,ghaffari2018derandomizing}.

\minusspace
\paragraph{The importance of the distributed LLL and sinkless orientation.}
In the LOCAL model, the distributed LLL and sinkless orientation have been of paramount importance for the development of entire lines of research in the LOCAL model.
Below we discuss a few of the highlights.

The aforementioned (randomized) lower bound of $\Omega(\log \log n)$ (which was lifted to a \emph{deterministic} lower bound of $\Omega(\log n)$ by~\cite{chang2019exponential}), together with matching randomized and deterministic upper bounds of $O(\log \log n)$ and $O(\log n)$, respectively, for sinkless orientation yielded the first proof that there exist locally checkable problems with provably \emph{intermediate} complexities on constant-degree graphs.
Following this discovery, a long line of research (see, e.g., \cite{brandt2017lcl,balliu2018new,balliu2019distributed,chang2019time,BalliuBEHMOS20,chang20complexity,rozhovn2020polylogarithmic,balliu2021almost,balliu2021locally,grunau2022landscape,chang24the}) mapped out the landscape of locally checkable problems on constant-degree graphs in a variety of settings, amongst others culminating in an almost complete understanding of the possible complexities that such problems can exhibit on trees and general (constant-degree) graphs.
One particularly intriguing result in this line of research is a randomized speedup result by~\cite{chang2019time}: in this work, the authors show that any problem in the aforementioned setting that admits a sublogarithmic-time algorithm can be solved as fast as the distributed LLL (with a polynomial LLL criterion).
This result implies a gap in the randomized complexity landscape which future work may be able to widen further by proving better upper bounds on the complexity of the distributed LLL.

However, the perhaps most important consequence of the aforementioned lower bound proved for the distributed LLL and sinkless orientation has been the development of a new lower bound technique in the LOCAL model called \emph{round elimination}.
Building on the lower bound proof in~\cite{brandt2016lower} which uses a specific version of round elimination fine-tuned to sinkless orientation, \cite{brandt2019automatic} generalized the used approach to a generic lower bound technique that, in principle, is applicable to any locally checkable problem.
Using this general version of round elimination, a variety of lower bounds for some of the most fundamental problems studied in the LOCAL model have been obtained~\cite{chang2019distributed,brandt2020truly,balliu2021lower,balliu2022distributed,balliu2022delta,balliu2023distributed}, emphasizing the importance of developing generic techniques.
Round elimination has even been used to prove a gap in the aforementioned complexity landscape~\cite{grunau2022landscape} and to prove a lower bound in the LOCAL model that is instrumental for showing quantum advantage for a locally checkable problem~\cite{balliu2025distributed}.

\minusspace
\paragraph{Lower bounds for the distributed LLL and sinkless orientation in quantum-LOCAL.}
In contrast to the lower bound situation in the LOCAL model, in the quantum-LOCAL model (and any stronger model), no super-constant lower bounds for the distributed LLL or sinkless orientation were known prior to our work.
In fact, even obtaining \emph{large constant} lower bounds for sinkless orientation (and therefore also the distributed LLL) in quantum-LOCAL and stronger models seemed out of reach of current techniques and was stated as an open problem in~\cite{suomela2024open}.
We obtain the following fundamental open problem.

\vspace{0.05cm}
\begin{tcolorbox}
    \begin{oq}\label{op2}
        Prove superconstant lower bounds for the distributed LLL and sinkless orientation in the quantum-LOCAL model (and stronger models).
    \end{oq}
\end{tcolorbox}
\vspace{0.05cm}

Solving~\Cref{op2} would also make significant progress on the more long-term question of fully determining the complexity of sinkless orientation in models related to quantum-LOCAL such as the non-signaling model~\cite[Question 7]{d2025limits}.

\subsection{Our Contributions}\label{sec:contributions}
We prove the following lower bound result in the strongest model discussed above, the randomized online-LOCAL model.
For a formal introduction to the randomized online-LOCAL model, we refer the reader to~\Cref{sec:approach}.
\begin{restatable}{theorem}{solb}\label{thm:solb}
	The complexity of sinkless orientation in the randomized online-LOCAL model is in $2^{\Omega(\log^* n)}$.
	Moreover, this lower bound holds already on trees with maximum degree $4$ in which no node has degree $3$.
\end{restatable}

As discussed above, sinkless orientation is a special case of the algorithmic LLL for the class of trees mentioned in~\Cref{thm:solb}.
Hence, the lower bound from~\Cref{thm:solb} also applies to the distributed LLL.
Moreover, the lower bounds for sinkless orientation and the distributed LLL naturally apply also in any model that is at least as weak as the randomized online-LOCAL model, thereby providing lower bounds across models studied in different fields.
Recalling the model overview in~\cite[Fig.\ 10, arxiv version]{akbari2025online}, we can summarize the concrete implications of~\Cref{thm:solb} as follows.
\begin{corollary}\label{cor:lllandso}
	The complexities of sinkless orientation and the distributed LLL (even with exponential LLL criterion $p \cdot 2^d \leq 1$) are in $2^{\Omega(\log^* n)}$, in any model that is at least as weak as the randomized online-LOCAL model, which includes, e.g., the non-signaling model, the bounded-dependence model, and the deterministic dynamic-LOCAL model.
	In particular, the complexity of the distributed LLL in the quantum-LOCAL model is in $2^{\Omega(\log^* n)}$.
\end{corollary}

This bound constitutes the first superconstant lower bound for the distributed LLL and sinkless orientation in any of the models mentioned in~\Cref{cor:lllandso} and in particular solves~\Cref{op2}.
Moreover, it makes substantial first progress on~\cite[Question 7]{d2025limits}.

While $2^{\Omega(\log^* n)}$ might appear as a strange complexity, we would not be surprised if our lower bound is actually asymptotically tight (in the exponent) for sinkless orientation in the randomized and deterministic online-LOCAL models and perhaps also beyond.
On a high level, the reason for this is that the way in which our lower bound instances are constructed appear to not be wasteful at all (asymptotically speaking) w.r.t.\ the number of nodes used in the construction, and the bound of $2^{\Omega(\log^* n)}$ naturally emerges from it via the fact that if $n$ nodes are required in our construction to prove a lower bound of $k$, then exponential in $n$ many nodes are required in our construction to prove a lower bound of $2k$.

\minusspace
\paragraph{Towards a generic post-quantum lower bound technique?}
The approach for proving our lower bound is based on an entirely new lower bound technique we develop in this work.
While the proof requires a lot of technical details, the actual graph constructions we use are based on conceptually simple graph operations, which provide an easy starting point for extensions and generalizations.
This is somewhat similar to the lower bound for sinkless orientation and the distributed LLL proved in the LOCAL model~\cite{brandt2016lower}, which, due to its \emph{conceptual} simplicity gave rise to the generic lower bound technique of round elimination.
In fact, there is another example of a similar historical behavior: a (conceptually elegant) lower bound technique from descriptive combinatorics that turned out to be applicable also to the LOCAL model, called Marks' technique~\cite{marks2016determinacy}, was essentially\footnote{Formally, the problem considered in~\cite{marks2016determinacy} is $\Delta$-coloring, to which there exists a simple reduction from sinkless orientation in the considered setting. This close connection also ensures that the same approach considered in that paper works directly for sinkless orientation.} developed in the context of sinkless orientation and has been subject to generalizations later~\cite{BCGGRV22,brandt2024homomorphism}.
In general, it seems that finding a conceptually simple lower bound approach for sinkless orientation is a highly promising first step towards providing a much more widely applicable lower bound technique in the respectively considered model and thereby towards solving~\Cref{op1}.
We believe that our lower bound approach constitutes such an approach and hope that it will serve as a foundation for many more post-quantum lower bounds for other locally checkable problems.

\subsection{Our Approach}\label{sec:approach}
In this section, we give a bird's eye view of our approach for proving~\Cref{thm:solb}.
As, at this point, the details of the considered model become relevant, we finally introduce the randomized online-LOCAL model formally.

\minusspace
\paragraph{The randomized online-LOCAL model.}
For simplicity, we first describe the deterministic online-LOCAL model, and then explain the minor changes required to obtain the randomized online-LOCAL model.
In the deterministic online-LOCAL model, different from the setting of the LOCAL model, the nodes of the input graph $G$ are presented to the algorithm one after another in an adversarial order.
Each time a node $v$ is presented, the (deterministic) algorithm obtains full information about the $L$-hop neighborhood of $v$, for some parameter $L$ (that, if considered for a class of graphs instead of for a single graph, is a function $L(n)$ of the number $n$ of nodes of the input graphs).
More specifically, the algorithm obtains complete information about the topology of the $L$-hop neighborhood of $v$ and about the degree of (and the port numbers at) each node in this neighborhood (including those at distance precisely $L$), and for each node $w$ in this neighborhood it receives the information which nodes that it has seen previously are identical to $w$.\footnote{One can imagine that the algorithm labels each node it sees with a unique identifier and can recognize those identifiers when it sees such a node again. However, we emphasize that, in contrast to the LOCAL model, the deterministic (and also the randomized) online-LOCAL model are formally defined without identifiers.}
Upon receiving this information, the algorithm has to decide on the output for $v$ before the next node and its $L$-hop neighborhood are revealed.
The algorithm is correct if the global output induced by the choices for each node is a correct solution to the considered problem.
We emphasize that the algorithm is not memory-constrained: it can remember all information that it receives during the execution.
The complexity (or locality) of such an algorithm is the aforementioned function $L(n)$ (w.r.t.\ worst-case input instances consisting of an input graph and an ordering of the nodes), i.e., the parameter indicating the radius of the neighborhood around a presented node that is given to the algorithm.

In the randomized online-LOCAL model, the algorithm has an unlimited source of randomness it can use for making its decisions.
Moreover, the adversary is oblivious, i.e., it has to select the input instance before the algorithm generates its random bits.
Furthermore, an algorithm is considered to solve a given problem if it produces a correct solution w.h.p., i.e., with probability at least $1 - 1/n$, on any input instance.
Consequently, the complexity of a problem in the randomized online-LOCAL model is the complexity of an optimal algorithm that produces a correct solution w.h.p.

For readers familiar with the SLOCAL model~\cite{ghaffari2017complexity}, we remark that the (deterministic and randomized) online-LOCAL models can be seen as versions of the (deterministic, resp.\ randomized) SLOCAL model with global memory.
Moreover, for discussing an online-LOCAL algorithm, it will be convenient to be able to refer to the nodes and edges the algorithm has already ``seen'' at some point of the execution.
\begin{definition}[seeing a node/edge]\label{def:onlinelocalmodel}
    Let $L$ denote the complexity of a (deterministic or randomized) online-LOCAL algorithm $\fA$ that is executed on some graph $G$.
    Consider any point of the execution, and let $v$ and $e = \{ u, u' \}$ denote a node and an edge of $G$, respectively.
    We say that $\fA$ \emph{has seen $v$} if there is a node $w \in V(G)$ that has been presented to $\fA$ (up to the considered point of the execution) and has distance at most $L$ from $v$.
    We say that $\fA$ \emph{has seen $e$} if there is a node $w \in V(G)$ that has been presented to $\fA$ and has distance at most $L$ from both $u$ and $u'$.
\end{definition}

Moreover, we will use the terms \emph{query} and \emph{query sequence} to refer to the presentation of a node by the adversary and the sequence of presented nodes, respectively. 
Now we are set to present an informal overview of our approach.

\minusspace
\paragraph{High-level overview.}
To obtain the lower bound from~\Cref{thm:solb}, intuitively, we first prove the lower bound in the \emph{deterministic} online-LOCAL model and then adapt the construction to also work in the \emph{randomized} online-LOCAL model.
While we will not explicitly refer to the deterministic online-LOCAL model in the formal proof provided in~\Cref{sec:ingredients,sec:lowerbound}, we will present the high-level overview here by following the aforementioned intuition.
We remark that the lifting of the deterministic lower bound to the randomized lower bound can also be achieved in a simple black-box fashion by applying~\cite[Lemma 7.6, arxiv version]{akbari2025online}; we hope that our white-box approach will be useful for future applications of the developed technique by yielding additional insights.

\begin{figure}[ht!]
    \vspace{1cm}
	\centering
	\includegraphics[scale=1]{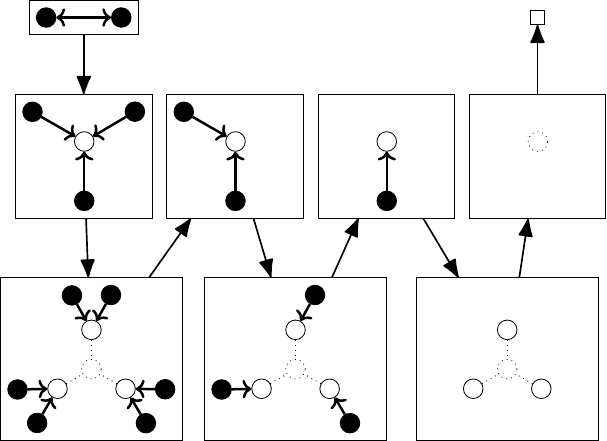}
	\caption{A high-level overview of our approach in the deterministic online-LOCAL model for the case $L = 2$ and $\Delta = 3$. Each box displays a configuration as discussed in the high-level overview in~\Cref{sec:approach}. A downward arrow between configuration corresponds to a reflection operation, an upward arrow to a split operation, where in the source configuration the node to be removed is indicated by a dotted circle and in the target configuration only one of the isomorphic connected subconfigurations of the resulting configuration is displayed. The final (small) box represents the empty configuration.}
	\label{fig:initialexample}
\end{figure}

Now imagine we want to prove a lower bound of $L$ for the complexity of sinkless orientation in the \emph{deterministic} online-LOCAL model.
Obtaining such a lower bound is equivalent to showing that any algorithm with complexity at most $L - 1$ fails on some input graph.
Hence, assume for a contradiction that there exists some algorithm $\fA$ solving sinkless orientation with a complexity of at most $L - 1$.
We will explicitly construct some input instance on which $\fA$ fails.
For an illustration of the initial part of our argumentation see~\Cref{fig:initialexample}, which illustrates the case $L = 2$ and $\Delta = 3$ (in which $\fA$ has complexity at most $L - 1 = 1$).

\minusspace
\paragraph{A sequence of contradictory configurations.}
In essence, our argumentation starts from an (imagined) incorrect local output and then works its way backwards to how $\fA$ can be actually forced to produce such an incorrect output.
Our starting point is a configuration representing the simplest possible incorrect output: two adjacent nodes that both choose the connecting edge to be oriented outwards.
However, why would the node $v$ that is presented to $\fA$ later than the other of the two not choose to orient the connecting edge inwards (thereby agreeing with the decision of the adjacent node) and orient another incident edge outwards?
Well, imagine that all neighbors of $v$ were presented to $\fA$ before $v$ and all of them chose to orient the edge connecting them to $v$ towards $v$.
This is a new configuration that enforces that the aforementioned contradictory configuration is produced by $\fA$, no matter which incident edge $v$ chooses to orient outwards.

In the same way as we obtained a new contradictory configuration from the initial one, we can now continue to iteratively transform an obtained configuration $C$ into a new configuration $C'$ that forces $\fA$ to produce $C$.
To guarantee that $\fA$ cannot avoid to produce $C$ when presented with a suitable node in $C'$, we create $C'$ from $C$ in the following way: we choose a node $u$ that, in $C$, is a leaf node that has chosen to output its incident edge away from $u$ and then consider as $C'$ the configuration obtained from $C$ by first undoing $u$'s output decision (which, in~\Cref{fig:initialexample}, corresponds to turning the edge $u$ had chosen to orient outwards into an undirected edge) and then taking $\Delta$ copies of $C$ and identifying all copies of $u$ in those copies.
Here $\Delta$ is some parameter that will correspond to the maximum degree of the graph obtained at the end of our construction.
Now, when presented with $u$ in $C'$, Algorithm $\fA$ cannot avoid to produce $C$.

\minusspace
\paragraph{Reflection and split operations.}
By the described operation---which we call the \emph{reflection operation}---, we can transform a contradictory configuration into a new one.
However, across a sequence of configurations produced by applying the reflection operation, the number of nodes that have already decided on their output grows instead of decreases while, ideally, we would like them to reach $0$ so that we can take the obtained graph as our input graph and the sequence of the nodes at the center of the reflection operations as query sequence (in reverse order).
To this end, we introduce another operation, called the \emph{split operation}, which essentially consists of removing a node from a considered configuration $C$ that is farther away from any node in $C$ that has already decided on its output than the complexity of $\fA$.
The justification for this removal is that, due to the mentioned distance property of the removed node, Algorithm $\fA$ cannot have seen the removed node, which implies that the behavior of $\fA$ is independent of whether we include the node or remove it.
Moreover, the removal disconnects the considered configuration into connected subconfigurations that (again due to the aforementioned distance property) are ``independent'' of each other w.r.t.\ the behavior of $\fA$; hence we can iterate on each connected subconfiguration separately.
Due to the symmetry generated by the reflection operation, by choosing the node to remove (amongst those nodes that satisfy the distance property) suitably, we can ensure that the removal performed by the split operation will result in isomorphic connected subconfiguration, which allows us to reduce attention to a single of those subconfigurations, implicitly assuming that whatever further operations we perform on that subconfiguration, we also perform on all isomorphic subconfigurations. 

\begin{figure}[ht!]
	\centering
	\includegraphics[scale=1]{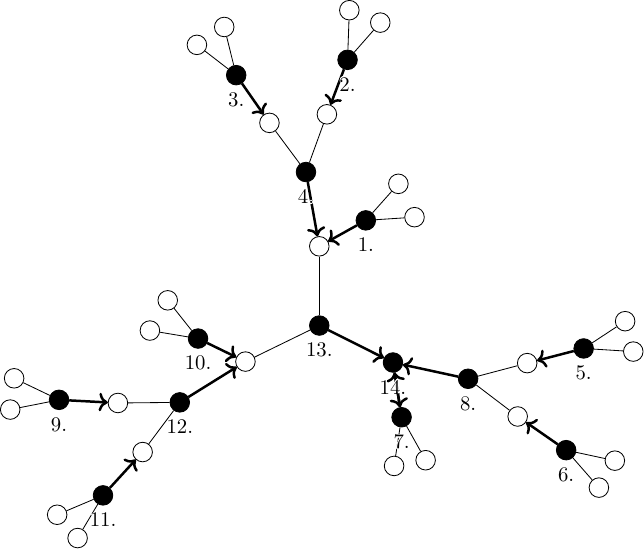}
	\caption{A slightly simplified overview of the input graph and query sequence obtained from the approach illustrated in~\Cref{fig:initialexample} for $L = 2$ (i.e., considering an algorithm $\fA$ with complexity $L - 1 = 1$) and $\Delta = 3$. The solid nodes represent the first nodes in the query sequence, in the order displayed below the nodes. For each solid node one incident edge is displayed as being oriented away from the node, illustrating the part of the decision of $\fA$ for the respective node that is relevant for obtaining the desired contradiction. Note that each time $\fA$ has to make a decision for a solid node $v$, the situation regarding the connected component containing $v$ of the subgraph that $\fA$ has already seen at this point in time is symmetric, essentially allowing us to choose which incident edge we want $\fA$ to orient outwards. With the $14$th presented node, $\fA$ will create an incorrect output.
    We remark that, unfortunately, the small examples given in this figure and~\Cref{fig:initialexample} are misleading in their simplicity; some assumptions that one is tempted to make by looking at the figures do actually not hold when considering larger $L$. For instance, the aforementioned symmetry does not hold in general; however it holds if one ignores the parts of the connected component that are ``behind'' a node that is oriented towards the considered node $v$. We also note that the aforementioned ``slight simplification'' of~\Cref{fig:initialqueryseq} merely consists in omitting some nodes that are incident to the displayed nodes and guarantee that every displayed node has degree $\Delta = 3$ (which is relevant due to the precise definition of the deterministic online-LOCAL model, in particular regarding which information the algorithm receives).}
	\label{fig:initialqueryseq}
\end{figure}

\minusspace
\paragraph{Deriving an input instance.}
Now imagine a sequence of reflection and split operations that ends up with an empty graph.
By reversing the sequence, i.e., by starting with the empty graph and traversing the sequence of reflection and split operations in reverse order, where, for a split operation, we \emph{add} the removed node and, for a reflection operation, present the node at the center of the operation to $\fA$, we obtain both a graph and a query sequence consisting of (a subset of) the nodes of the obtained graph.
This graph and this query sequence will now essentially constitute our input instance for which $\fA$ will fail.
The reason why $\fA$ fails on this instance is that the construction ensures that $\fA$ produces precisely the contradictory configurations discussed above in reverse order until it reaches the incorrect output of the initial configuration.\footnote{There is a hidden subtlety in our argumentation: So far, we have implicitly assumed that we can decide \emph{which} edge $\fA$ orients outwards from a node that is queried, based on the fact that \emph{any} decision will lead to the previous configuration in the above sequence of configurations. In principle, while, for each considered step, $\fA$ cannot avoid producing the configuration we desire, it can choose a different edge to orient outwards than we intend in the overall construction. However, this freedom of choice of $\fA$ is of cosmetic nature: due to the local isomorphisms in our construction, we can essentially continue the construction of our input instance in a locally isomorphic manner (compared to the presented fixed construction), no matter which edge $\fA$ chooses.}
The input instance derived for the case $L = 2$ and $\Delta = 3$ corresponding to~\Cref{fig:initialexample} is illustrated in~\Cref{fig:initialqueryseq}.

\minusspace
\paragraph{The construction tree.}
On a high level, the above discussion reduces the task of proving the desired lower bound in the deterministic online-LOCAL model to the task of finding a suitable sequence of reflection and split operations that results in the empty graph.
Note that the applicability of the split operation (which we would like to apply as often as possible to reduce the size of the considered graph(s)) is limited due to the required distance property.
In fact, it is far from clear that a sequence as discussed above resulting in the empty graph even exists.

Our construction of such a sequence is conceptually simple although there are many technical details that have to be taken care of.
The key object that we use to obtain such a sequence is what we call a \emph{construction tree}.
A construction tree encodes which (reflection/split) operations to perform at which nodes in which order.
More precisely, the definition of a construction tree already ensures that the sequence of operations that can be read off of the construction tree is ``good'' in various regards.
Amongst others, the construction tree satisfies a number of properties related to guaranteeing that the graph obtained at the end of the sequence of operations is empty and that the sequence of graphs obtained during the sequence of operations has lots of symmetries.
(We remark that, for technical reasons, in our proof we do not remove nodes when performing a split operation, but instead \emph{mark} them; hence, formally, the guarantee that the obtained graph is empty translates to the guarantee that all nodes of the obtained graph are marked.)

\minusspace
\paragraph{Deriving the lower bound in the deterministic online-LOCAL model.}
Perhaps most importantly, based on the structure of the construction tree, we can define a transformation $F(\cdot)$ that takes a construction tree as input and returns a graph that can be shown to be another (usually much larger) construction tree.
This transformation is carefully designed to satisfy a key property that lies at the heart of our lower bound approach: if $T$ is a construction tree that yields a sequence of reflection and split operations that results in the empty (respectively fully marked) graph and obeys the aforementioned distance property with parameter $L - 1$ (describing the complexity of the considered algorithm $\fA$), then $F(T)$ is a construction tree that yields a sequence that results in the empty graph and obeys the aforementioned distance property with parameter $2L - 1$.
In other words, if the lower bound that we obtain by using construction tree $T$ is $L$, then the lower bound that we obtain by using construction tree $F(T)$ is $2L$.

Moreover, we show that if we compare the number of nodes of the input graph that we obtain by using construction tree $T$ with the number of nodes of the input graph that we obtain by using construction tree $F(T)$, then the latter is exponential in the former.
In other words, by increasing the size of the input graph exponentially, we increase the obtained lower bound by a factor of $2$, which results in the desired lower bound of $2^{\Omega(\log^* n)}$ (though so far only for the \emph{deterministic} online-LOCAL model).  

\minusspace
\paragraph{Lifting the lower bound to the randomized online-LOCAL model.}
In order to lift the obtained lower bound to the \emph{randomized} setting, we modify, for each deterministic lower bound instance, both the input graph and the query sequence by a sequence of adaptations each of which corresponds to a presented node.
More specifically, consider any of the lower bound instances constructed for the deterministic online-LOCAL model.
We essentially transform this instance into a lower bound instance for the randomized online-LOCAL model by going through the query sequence and, each time a node is presented to the algorithm, tweaking both the overall graph and the remainder of the query sequence in a manner depending on the \emph{most probable} behavior of the considered randomized algorithm w.r.t. which edge(s) incident to the presented node the algorithm will orient outwards.
In essence, these modifications guarantee that we obtain the same sequence of contradicting configurations as in the deterministic lower bound proof until we reach an incorrect output, with the difference that for each presented node we may now lose a factor of $\Delta$ in the probability that the behavior is indeed as we need for our lower bound.
(Note that we do not lose more than a factor of $\Delta$ due to considering the ``most probable'' behavior of the considered randomized algorithm, which occurs with probability at least $1/\Delta$.)

While overall, this results in our contradiction occurring with only a small probability, by increasing the number $n$ of nodes suitably (by adding sufficiently many nodes to the considered input graph) we can make sure that this small probability is still larger than $1/n$, thereby guaranteeing that the considered algorithm fails with probability more than $1/n$, yielding a contradiction.
For this, an exponential increase in the number of nodes suffices, which yields a lower bound of $2^{\Omega(\log^*(\log n))}$ which is identical to $2^{\Omega(\log^* n)}$.
Hence, asymptotically (in the exponent), we obtain the same lower bound for the randomized online-LOCAL model as for the deterministic online-LOCAL model.

\subsection{Roadmap}
In~\Cref{sec:ingredients}, we collect a number of ingredients on which our lower bound proof builds.
More specifically, \Cref{sec:ingredients} contains the following parts.
In~\Cref{sec:constructiontree} we develop the notion of a construction tree.
In~\Cref{sec:seq} we make use of this general definition to define a concrete sequence of construction trees that we will use to derive our lower bound proof.
To obtain this sequence, we introduce a graph transformation and prove that it preserves the property of being a construction tree.
We also introduce the ``initial tree'' from which we obtain the aforementioned sequence via iterative application of the aforementioned transformation and bound the number of nodes of the trees in the sequence.
In~\Cref{sec:aux}, we define two edge labelings for construction trees that, in particular, will be useful for keeping track of the different components obtained from the removal of nodes discussed in the above high-level overview.
Moreover, we prove a number of technical lemmas about these labelings.
In~\Cref{sec:inputtree}, we define an input tree that is already very close to the input graph that we will present to the algorithm in~\Cref{sec:lowerbound} and prove its well-definedness.
In~\Cref{sec:relationtft}, we introduce the notion of distance-$D$-correctness and prove the key lemma that if a construction tree $T$ is distance-$D$-correct, then the tree obtained from $T$ by the aforementioned graph transformation is distance-$(2D)$-correct.
Since the property that a construction tree is distance-$D$-correct roughly corresponds to the property that the input instance obtained from the construction tree requires complexity $D$ to be solved (in the deterministic online-LOCAL model), this key lemma, together with the bound on the size of the construction trees discussed in~\Cref{sec:seq}, provides the basis for the analysis of the complexity lower bound.
Finally, in~\Cref{sec:lowerbound}, we prove our lower bound by combining the above ingredients with new ingredients that are specifically targeted towards obtaining lower bound instances that are good against \emph{randomized} online-LOCAL algorithms.

\section{Ingredients for the Lower Bound Construction}\label{sec:ingredients}
In this section, we will develop a variety of tools and constructions that are essential for our overall lower bound approach.
Besides defining the objects we need in the approach, we will also prove a number of lemmas that collect important properties of said objects.

We start by recalling some basic graph-theoretic definitions and notations.
\begin{definition}
    For a graph $G$, we denote the set of nodes of $G$ by $V(G)$ and the set of edges of $G$ by $E(G)$, and write $G = (V(G), E(G))$.
    A \emph{(maximal) connected component} $C$ of $G$ is a connected subgraph of $G$ such that, for any two nodes $v \in V(C), w \in V(G) \setminus V(C)$, there is no path in $G$ connecting $v$ and $w$.
    Let $V' \subseteq V(G)$ be a subset of the nodes of $G$.
    The (sub)graph \emph{induced by $V'$} is the subgraph of $G$ with node set $V'$ in which any two nodes of $V'$ are connected if and only if they are connected in $G$.
    
    For two nodes $v, w$ of a rooted tree $T$, we call $w$ a \emph{descendant} of $v$ (and equivalently $v$ an \emph{ancestor} of $w$) if there is a directed path from $w$ to $v$ in $T$.
    This path may be of length $0$, i.e., any node is descendant and ancestor of itself.
    If there is neither a directed path from $v$ to $w$, nor a directed path from $w$ to $v$, then we call $v$ and $w$ \emph{incomparable}.
    A \emph{leaf (node)} of a (rooted) tree is a node of degree $1$.
    For a directed edge $e = (v, w)$, we call $v$ the \emph{tail} of $e$ and $w$ the \emph{head} of $e$.

    In order to distinguish between the two ports numbers assigned to the same edge $e$ with endpoints $v$ and $w$, we will refer to them as the \emph{port number assigned by $v$ (resp.\ $w$) to $e$} or simply as the \emph{port number at $v$ (resp.\ $w$)}.
\end{definition}

\subsection{The Construction Tree}\label{sec:constructiontree}
In this section, we define one of the crucial building blocks of our approach, which we call a \emph{construction tree}.
A construction tree can be thought of as a rooted tree with certain properties that, roughly speaking, encodes how to construct a hard lower bound instance, regarding both the input graph and the query sequence (that will force an algorithm to have at least a certain complexity).
Please note that this high-level description is somewhat inaccurate in that the precise construction of a lower bound instance will depend on the considered algorithm; however, such a lower bound instance for a concrete algorithm will be achieved by tweaking the aforementioned general ``hard lower bound instance''.

Before we can introduce construction trees formally, we need to introduce a number of additional definitions.

\minusspace
\paragraph{Topological considerations.}
We first collect some definitions that are relevant for defining the topological structure of a construction tree.
The following definition introduces a balance property that construction trees will satisfy. 

\begin{definition}[$b$-balanced tree]\label{def:bbalanced}
    Let $T$ be a rooted tree with root $r$.
    For each positive integer $i$, we denote the set of vertices of $T$ that are at distance precisely $i - 1$ of $r$ by $L_i$ (and will call such a set a \emph{layer}).
    For instance, $L_1 = \{ r \}$.
    
    Now, let $b \geq 3$ be an integer.
    We call $T$ \emph{$b$-balanced} if
    \begin{enumerate}
        \item every node of $T$ has precisely $0$, $1$ or $b$ children, and
        \item for each positive integer $i$ and any two nodes $v, w \in L_i$, the number of children of $v$ and $w$ is identical.  
    \end{enumerate}  
    We call a node a \emph{reflect node} if it has precisely one child, and a \emph{split node} otherwise.
    We denote the set of all reflect nodes by $V_R$ and the set of all split nodes by $V_S$.
    Moreover, we call a split node an \emph{internal split node} if it has precisely $b$ children, and a \emph{leaf (split) node} otherwise (i.e., if it has no children).

    We call a layer $L_i$ in a $b$-balanced rooted tree a \emph{reflect layer} if it consists only of reflect nodes and a \emph{split layer} if it consists only of split nodes.
    Moreover, we call $L_i$ an \emph{internal split layer} if it consists only of internal split nodes and a \emph{leaf (split) layer} if it consists only of leaf split nodes.
\end{definition}

We observe that for each internal split node of a $b$-balanced rooted tree, the $b$ subtrees hanging from its $b$ children are isomorphic rooted trees.
Moreover, it follows from the definition of a $b$-balanced rooted tree that each nonempty layer of a $b$-balanced rooted tree is either a reflect layer or a split layer, and each split layer is either an internal split layer or a leaf split layer (of which there is precisely one).
For simplicity we will disregard empty layers in the remainder of the paper, considering a layer $L_i$ to be a layer of a $b$-balanced rooted tree if and only if it is nonempty.

Next, we introduce a second property that construction trees will satisfy.
This property ensures that the layers from~\Cref{def:bbalanced} are ``well-nested'', regarding whether they are reflect or split layers.

\begin{definition}[well-nested]\label{def:wellnested}
    Let $T$ be a $b$-balanced rooted tree $T$ and let $i_{\max}$ denote the index of the leaf split layer (i.e., the layers of $T$ are precisely $L_1, \dots, L_{i_{\max}}$).
    We denote the set of indices of reflect layers by $\fI_{\text{reflect}}$ (i.e., $\fI_{\text{reflect}} := \{ i \mid 1 \leq i \leq i_{\max}, L_i \text{ is a reflect layer} \}$) and the set of indices of split layers by $\fI_{\text{split}}$ (i.e., $\fI_{\text{split}} := \{ i \mid 1 \leq i \leq i_{\max}, L_i \text{ is a split layer} \}$).
    We call $T$ \emph{well-nested} if
    \begin{enumerate}
        \item the number of split layers of $T$ equals the number of reflect layers of $T$,
        \item $1 \in \fI_{\text{reflect}}$ and $i_{\max} \in \fI_{\text{split}}$, and
        \item there is a bijection $\varphi: \fI_{\text{reflect}} \rightarrow \fI_{\text{split}}$ such that
        \begin{enumerate}
            \item $\varphi(1) = i_{\max}$,
            \item for each $i \in \fI_{\text{reflect}}$, we have $i < \varphi(i)$, and
            \item \label{item:nest} for each $i, j \in \fI_{\text{reflect}}$ with $i < j < \varphi(i)$, we have $\varphi(i) > \varphi(j)$.
        \end{enumerate}
    \end{enumerate}
\end{definition}

The following lemma guarantees that the bijection $\varphi$ from~\Cref{def:wellnested} is unique.

\begin{lemma}\label{lem:uniquebijection}
    Let $T$ be a well-nested $b$-balanced rooted tree.
    Then there is precisely one bijection $\varphi$ as described in \Cref{def:wellnested}.
\end{lemma}
\begin{proof}
    Let $\varphi$ be an arbitrary bijection with the properties described in \Cref{def:wellnested}.
    Consider some arbitrary $i \in \fI_{\text{reflect}}$.
    Let $j_{\min} \in \{ i + 1, \dots, i_{\max}\}$ denote the smallest index $j$ satisfying that the set $\{ L_i \dots, L_j \}$ contains equally many reflect and split layers.
    We note that $j_{\min}$ exists and is contained in $\fI_{\text{split}}$, by the fact that $T$ is well-nested.
    Moreover, we claim that $\varphi(i)$ must be identical to $j_{\min}$.

    Assume for a contradiction that this is not the case.
    Consider first the case that $\varphi(i) > j_{\min}$.
    Then, by the properties of $\varphi$ described in \Cref{def:wellnested}, we have $i < \varphi^{-1}(j_{\min}) < j_{\min}$.
    If the set $\{ L_{\varphi^{-1}(j_{\min})}, \dots, L_{j_{\min}} \}$ contains equally many reflect and split layers, then so does the set $\{ L_i, \dots, L_{\varphi^{-1}(j_{\min}) - 1} \}$, contradicting the minimality of $j_{\min}$.
    Hence, assume that $\{ L_{\varphi^{-1}(j_{\min})}, \dots, L_{j_{\min}} \}$ does not contain equally many reflect and split layers.
    If it contains more reflect layers than split layers, then by the pigeonhole principle there must be some index $\varphi^{-1}(j_{\min}) < k < j_{\min}$ of a reflect layer satisfying $\varphi(k) > j_{\min}$, violating Property~\ref{item:nest} in \Cref{def:wellnested} for $\varphi^{-1}(j_{\min})$ and $k$.
    If $\{ L_{\varphi^{-1}(j_{\min})}, \dots, L_{j_{\min}} \}$ contains more split layers than reflect layers, then, similarly, there must be some index $\varphi^{-1}(j_{\min}) < k < j_{\min}$ of a split layer satisfying $\varphi^{-1}(k) < \varphi^{-1}(j_{\min})$, violating Property~\ref{item:nest} in \Cref{def:wellnested} for $\varphi^{-1}(k)$ and $\varphi^{-1}(j_{\min})$.

    Now consider the second case, i.e., that $\varphi(i) < j_{\min}$.
    Then, analogously to above, we obtain that the set $L_i, \dots, L_{\varphi(i)}$ does not contain equally many reflect and split layers (as otherwise the minimality of $j_{\min}$ would be violated).
    An analogous argumentation to above then yields that there is some $i < k < \varphi(i)$ satisfying $\varphi(i) < \varphi(k)$ or $\varphi^{-1}(k) < \varphi(i)$, yielding a contradiction to Property~\ref{item:nest} in \Cref{def:wellnested} in either case.
    This concludes the proof by contradiction and proves the claim that $\varphi(i) = j$.

    Since $i$ and $\varphi$ were chosen arbitrarily, we obtain that there is at most one bijection $\varphi$ as described in \Cref{def:wellnested}.
    As, by definition, there exists such a bijection due to the well-nestedness of $T$, it follows that there is exactly one such bijection $\varphi$.
\end{proof}

Now, \Cref{lem:uniquebijection} gives rise to the following definition which indexes reflect and split layers.

\begin{definition}[$j$-th reflect/split layer]
    Let $T$ be a well-nested $b$-balanced rooted tree.
    Let $\varphi$ be the unique bijection from \Cref{lem:uniquebijection}.
    Let $i_1 < i_2 < \dots < i_{|\fI_{\text{reflect}}|}$ be the elements of $\fI_{\text{reflect}}$, ordered by size.
    Then, for each $1 \leq j \leq |\fI_{\text{reflect}}|$, we set $\fR_j := L_{i_j}$.
    We call $\fR_j$ the \emph{$j$-th reflect layer (of $T$)}.
    Moreover, for each $1 \leq j \leq |\fI_{\text{reflect}}|$, we set $\fS_j := L_{\varphi(i_j)}$ and call $\fS_j$ the \emph{$j$-th split layer (of $T$)}.

    For any node $v$ in a layer $L_i$, we call $i$ the \emph{layer index} of $v$.
    Similarly, for any node $v$ in a reflect layer $\fR_j$ or split layer $\fS_k$, we call $j$ the \emph{reflect index} or $k$ the \emph{split index}, respectively, of $v$.
    If $L_i = \fR_j$, we call $i$ the \emph{layer index} of $\fR_j$ and $j$ the \emph{reflect index} of $L_i$.
    Similarly, if $L_i = \fS_k$, we call $i$ the \emph{layer index} of $\fS_k$ and $k$ the \emph{split index} of $L_i$.
\end{definition}

Note that, for indices $j < k$, the $j$-th split layer $\fS_j$ might be at greater depth from the root $r$ than the $k$-th split layer $\fS_k$ (whereas for reflect layers, $\fR_k$ is at greater depth from $r$ than $\fR_j$, for all $j < k$ for which $\fR_j$ and $\fR_k$ exist).

\minusspace
\paragraph{Label considerations.}
After the above topological considerations, we turn our attention now towards tree labelings since the definition of a construction tree will also involve node labels.
Our next definition collects a number of useful notions related to strings/labels.

\begin{definition}[final substring/independent/clearing]\label{def:indepclear}
    Let $b \geq 3$ be an integer.
    We denote by $X_b$ the set of all strings of the form $x_j x_{j - 1} \dots x_0$ where $j$ is a nonnegative integer, $x_0 \in \{ 1, 2 \}$, and, for all $1 \leq i \leq j$, we have $x_i \in \{ 1, 2, \dots, b, * \}$, i.e., $X_b = \{ x_j x_{j - 1} \dots x_0 \mid j \in \mathbb{Z}_{\geq 0}, x_0 \in \{ 1, 2 \}, x_i \in \{ 1, 2, \dots, b , * \} \text{ for all } 1 \leq i \leq j \}$.
    Please note that throughout the paper, we will index strings from $X_b$ with indices starting at $0$ (starting at the right-most symbol).
    Moreover, for simplicity, we will write the concatenation of strings and symbols in a compact way, e.g., for some string $x = x_j \dots x_0$ and some symbol $z$, we write $xz$ to denote the string $x_j \dots x_0 z$ and $zx$ to denote the string $z x_j \dots x_0$.
    
    Consider some string $x = x_j x_{j - 1} \dots x_0 \in X_b$.
    For each $0 \leq i \leq j$, we call $x_i$ the $i$-th symbol of $x$.
    In particular, $x_0$ is the $0$-th symbol of $x$.
    Moreover, we define the \emph{length} $\len(x)$ of $x$ to be $j + 1$.
    For two strings $x, x' \in X_b$ we call $x$ a \emph{final substring} of $x'$ if $\len(x) \leq \len(x')$ and $x_i = x'_i$ for each $0 \leq i \leq \len(x) - 1$.
    For a finite subset $Y$ of $X_b$, we define the length $\len(Y) := \max_{y \in Y} \len(y)$ of $Y$ to be the maximum length of a string contained in $Y$.
    Furthermore, we define $X_b^-$ as the subset of $X_b$ containing all strings that do not contain the symbol $*$ (i.e., that just use the symbols $1, \dots, b$).

    Now consider some finite subset $Y$ of $X_b$.
    We call $Y$ \emph{independent} if, for any two strings $y, y' \in Y$ satisfying $y \neq y'$, neither of $y$ and $y'$ is a final substring of the other.
    Finally, consider some finite subset $Y'$ of $X_b^-$.
    We call $Y'$ \emph{clearing} if, for any string $x \in X_b^-$ of length $\len(x) \geq \len(Y')$, there exists a string $y' \in Y'$ that is a final substring of $x$. 
\end{definition}

\begin{figure}[ht!]
	\centering
	\includegraphics[scale=1]{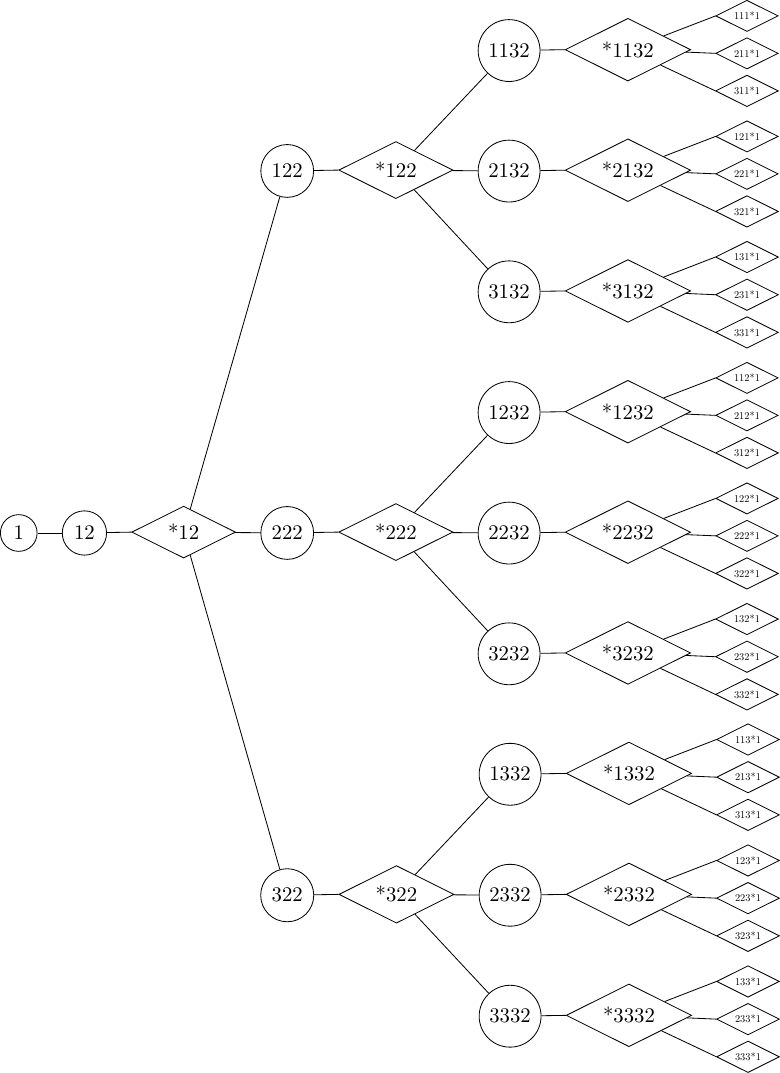}
	\caption{A $b$-ary construction tree for $b = 3$. Circles and diamonds represent reflect nodes and split nodes, respectively. The string inside a circle/diamond is the label of the respective node.}
	\label{fig:constructiontree}
\end{figure}

Now we are ready to finally define a construction tree.
Essentially, a construction tree is a well-nested balanced rooted tree with a node labeling that satisfies a list of carefully selected properties.
For an illustration of a construction tree, see~\Cref{fig:constructiontree}.

\begin{definition}[solid/construction tree]\label{def:solidity}
    Let $b \geq 3$ be an integer.
    Let $T$ be a well-nested $b$-balanced rooted tree with root $r$.
    Consider a labeling of the nodes of $T$ with labels from $X_b$, i.e., a function $\ell: V(T) \rightarrow X_b$.
    We call $\ell$ \emph{solid} if it satisfies the following properties.
    \begin{enumerate}
        \item\label{prop:rootone} $\ell(r) = 1$.
        \item\label{prop:reflectlabellength} For each $i \geq 1$ and each reflect node $v$ in reflect layer $\fR_i$, the length of label $\ell(v)$ is $i$.
        \item\label{prop:nosamereflect} For any two reflect nodes $v, w \in V_R$ with $v \neq w$, we have $\ell(v) \neq \ell(w)$.
        \item\label{prop:reflectnostar} For each reflect node $v$, label $\ell(v)$ does not contain the symbol $*$, i.e., $\ell(v) \in X_b^-$.
        \item\label{prop:indepandclearing} The set $\{ \ell(v) \mid v \in V_R \}$ of labels of all reflect nodes is independent and clearing.
        \item\label{prop:numberplusone} For each $i \geq 1$ and each split node $v$ in layer $L_i$, the length of label $\ell(v)$ is larger by precisely $1$ than the number of indices $1 \leq j < i$ such that $L_j$ is a reflect layer.
        \item\label{prop:splitstar} Consider any split node $v$. Let $i$ be the split index of $v$ and let $w$ be the unique ancestor of $v$ in reflect layer $\fR_i$. Then the $i$-th symbol of $\ell(v)$ is $*$ and the $0$-th to $(i-1)$-th symbols of $\ell(v)$ are the same as in $\ell(w)$.
        Moreover, $\ell(v)$ contains the symbol $*$ precisely once.
        \item\label{prop:splitkids} Consider any internal split node $v$ and let $i$ be the split index of $v$. Let $w_1, \dots, w_b$ denote the $b$ children of $v$. Then there exists a permutation $\rho: \{ 1, \dots, b \} \rightarrow \{ 1, \dots, b \}$ such that, for each $1 \leq j \leq b$ and each descendant $u$ of $w_j$ (including $w_j$ itself), the $i$-th symbol of $\ell(u)$ is $\rho(j)$.
    \end{enumerate}

    We call a pair $(T, \ell)$ a \emph{($b$-ary) construction tree} if $T$ is a well-nested $b$-balanced rooted tree and $\ell$ a solid labeling of $T$.
    For simplicity, we may omit $\ell$ and use $T$ instead of $(T, \ell)$.

    Moreover, for each internal split node $v$ in some layer $\fS_i$ and each $1 \leq j \leq b$, call the child $w_j$ of $v$ satisfying that the $i$-th symbol of $\ell(w_j)$ is $j$ the \emph{$j$-th child of $v$}.
    (Note that all those children exist due to Property~\ref{prop:splitkids}.)
\end{definition}

The following observation collects some simple properties of construction trees.

\begin{observation}\label{obs:basicconstree}
    Let $T$ be a construction tree. Then the following properties hold.
    \begin{enumerate}
        \item\label{obsitem:one} The root $r$ of $T$ is a reflect node; the leaves of $T$ are split nodes.
        \item\label{obsitem:two} The $0$-th symbol of the label of a node of $T$ is $1$ if and only if the node is the root or a leaf; the $0$-th symbol of the label of a node of $T$ is $2$ otherwise.
        \item\label{obsitem:three} For each node $v \in V(T)$, the length of $\ell(v)$ is larger by precisely $1$ than the number of reflect nodes on the path from $v$ to $r$ that are distinct from $v$.
    \end{enumerate}
\end{observation}
\begin{proof}
    Property~\ref{obsitem:one} follows from the well-nestedness of $T$.
    Property~\ref{obsitem:two} follows from the well-nestedness of $T$ and Properties~\ref{prop:rootone}, \ref{prop:indepandclearing}, and~\ref{prop:splitstar} in~\Cref{def:solidity}.
    Property~\ref{obsitem:three} follows from Properties~\ref{prop:reflectlabellength} and~\ref{prop:numberplusone} in~\Cref{def:solidity}.
\end{proof}

\subsection{A Sequence of Construction Trees}\label{sec:seq}
While~\Cref{sec:constructiontree} took care of defining the general concept of a construction tree, the goal of the current section is to define a sequence $T_2, T_3, \dots$ of infinitely many concrete construction trees (with a growing number of nodes) that we will use later for our lower bound construction.
For defining this sequence, we will develop a transformation $F(\cdot)$ that takes as input a construction tree and returns a construction tree.
Before defining $F(\cdot)$, we introduce the useful notion of a variant of a given string obtained by a certain padding operation.

\begin{definition}[padded version of a string]
    Let $b \geq 3$, $i \geq 1$, and $j \geq 0$ be integers. 
    Let $x \in X_b$ be a string of length $i$ and $\fK$ a subset of $\{ 0, \dots, i + j - 1 \}$ of size $|\fK| = j$.
    Let $p_0 < p_1 < \dots < p_{i - 1}$ denote the elements of $\{ 0, \dots, i + j - 1 \} \setminus \fK$.
    Then the \emph{$\fK$-padded version of $x$} is the string $y$ with symbols from the set $\{ 1, \dots, b, *, \square \}$ given by setting $y_{p_q} := x_q$, for each $0 \leq q \leq i - 1$, and $y_k := \square$, for each $k \in \fK$.
    In other words, $y$ is the string obtained from $x$ by spreading out the symbols of $x$ such that they occupy the positions whose index is not contained in $\fK$ (but in $\{ 0, \dots, i + j - 1\}$) and filling the positions whose index is contained in $\fK$ with the special symbol $\square$.
\end{definition}

Now we are set to define $F(\cdot)$.

\begin{definition}[transformation $F(\cdot)$]\label{def:fcdot}
    Let $T$ be a $b$-ary construction tree.
    Recall the definition of the $j$-th child of a split node in \Cref{def:solidity}.
    Consider the unique depth-first search traversal of $T$ where, for each split node and any $1 \leq i < j \leq b$, the $i$-th child of the split node is visited before the $j$-th child of the split node.
    Let $Q = (v_1, \dots, v_{2|V_R|})$ be the (chronologically ordered) sequence of reflect nodes visited by the traversal where each reflect node appears twice, once when it is visited for the first time (when the node is left towards its unique child) and once when it is visited for the second time (when the node is left towards its parent, unless it is the root).
    Note that \Cref{obs:basicconstree} ensures that no reflect node is a leaf of $T$, which implies that each reflect node is indeed visited twice.
    Furthermore, for simplicity, we may consider the same node occurring in two different places in $Q$ as two different objects.

    Call two nodes in $Q$ \emph{twins} if they correspond to the first and second appearance of the same node of $T$.
    Call a node in $Q$ \emph{early} if it is the first appearance of a node from $T$ in $Q$ and \emph{late} if it is the second appearance.
    If $v_i$ is the first appearance of a reflect node $u \in V(T)$ in $Q$ and $v_i$ is the $j$-th early node in $Q$, then we call $j$ the \emph{early rank} of both $u$ and $v_i$.
    Moreover, for each $1 \leq i \leq 2|V_R|$, let $\Lambda_i$ denote the number of late nodes in $\{ v_1, \dots, v_{i - 1} \}$.

    We now define $F(T)$ as follows.
    First we consider the topology of $F(T)$.
    We define the node set of $F(T)$ by $V(F(T)) := \bigcup_{1 \leq i \leq 2|V_R|} L^F_i$ where $L^F_i$ is a set of $b^{\Lambda_i}$ (so far isolated) nodes.
    To obtain the edge set $E(F(T))$ of $F(T)$, connect, for each $1 \leq i \leq 2|V_R| - 1$, each node in $L^F_i$ with \begin{enumerate}
        \item precisely one node in $L^F_{i + 1}$ if $v_i$ is an early node, and
        \item precisely $b$ nodes in $L^F_{i + 1}$ if $v_i$ is a late node,
    \end{enumerate} 
    in either case such that each node in $L^F_{i + 1}$ is connected to precisely one node in $L^F_i$ and the tail of the connecting edge is the respective endpoint in $L^F_{i + 1}$ and the head the respective endpoint in $L^F_i$.
    This is possible (and uniquely defines the topology of $F(T)$ up to isomorphism) as, by design, $|L^F_{i + 1}| = |L^F_i|$ if $v_i$ is an early node, and $|L^F_{i + 1}| = b \cdot |L^F_i|$ if $v_i$ is a late node.
    
    Observe that if $v_i$ is early, then each node in $L^F_i$ is a reflect node, and if $v_i$ is late, then each node in $L^F_i$ is a split node (for each $1 \leq i \leq 2|V_R|$).
    For each $1 \leq j \leq |V_R|$, set $\fR^F_j := L^F_i$ where $1 \leq i \leq 2|V_R|$ is the index such that $v_i$ has early rank $j$.
    Moreover, for each $1 \leq j \leq |V_R|$, set $\fS^F_j := L^F_{i'}$ where $1 \leq i' \leq 2|V_R|$ is the index such that, for the index $i$ satisfying that $v_i$ and $v_{i'}$ are twins, we have $\fR^F_j = L^F_i$.
    Note that the $\fR^F_j$ are precisely the reflect layers of $F(T)$ and the $\fS^F_j$ the split layers of $F(T)$.
    Note further that the fact that sequence $Q$ is obtained from a depth-first search traversal guarantees that $F(T)$ is a well-nested $b$-balanced rooted tree (which in particular implies that notions such as the split index of a node are well-defined in $F(T)$).

    For defining the labels of the nodes of $F(T)$, we start by defining, for each reflect layer and each split layer, the set of labels of the nodes in the respective layer.
    Consider any $1 \leq i \leq 2|V_R|$, and let $\fK \subseteq \{ 1, \dots, |V_R| \}$ be the set of indices $k$ such that $\fS^F_k$ has layer index $< i$.
    
    Consider first the case that $L^F_i$ is a reflect layer, and let $j$ be the reflect index of $L^F_i$.
    Then we define the labels of the nodes in $L^F_i$ to be precisely those strings in $X_b^-$ of length $j$ that can be obtained from the $\fK$-padded version of $\ell(v_i)$ by replacing the special symbols $\square$ with symbols from $\{ 1, \dots, b\}$ in all possible ways.
    (Note that, by the definitions of $Q$ and $\fK$ and Property~\ref{prop:reflectlabellength} of \Cref{def:solidity}, we have that $\fK$ is a subset of $\{ 0, \dots, |\fK| + \len(\ell(v_i)) - 1\}$, which implies that the $\fK$-padded version of $\ell(v_i)$ is well-defined.)
    For instance, for $b = 2$, $j = 5$, $K = \{ 2, 3\}$, and $\ell(v_i) = 211$, we obtain that the labels for the nodes in $L^F_i$ are $21111$, $21211$, $22111$, and $22211$.\footnote{This is just for the purpose of illustration; we do not claim that this case can actually occur in $F(T)$.}

    Now consider the case that $L^F_i$ is a split layer, and let
    $j$ be the number of reflect layers with a strictly smaller layer index than $L^F_i$.
    Then we define the labels of the nodes in $L^F_i$ to be precisely those strings in $X_b$ of length $j + 1$
    that can be obtained from the $\fK$-padded version of $*\ell(v_i)$ by replacing the special symbols $\square$ with symbols from $\{ 1, \dots, b\}$ in all possible ways.

    Note that, in either case, the size of the obtained set of labels is precisely the number of nodes in $L^F_i$ since there are precisely $|\fK|$ split layers with layer index $<i$ (which implies that the number of nodes in $L^F_i$ is $b^{|\fK|}$).
    In either case, we will assign each of the labels to precisely one node in $L^F_i$.
    It remains to decide for each layer which node is assigned which of the labels.

    We will do so in a way that satisfies the following property.
    Let $v \in L^F_i$ be a node of $F(T)$ (where $1 \leq i \leq 2|V_R| - 1$) and let $\fK$ be defined as above.
    Then, if $v$ is a reflect node, it is connected to the unique node $w \in L^F_{i + 1}$ satisfying $\ell(v)_p = \ell(w)_p$ for each $p \in \fK$; if $v$ is a split node, it is connected to the $b$ nodes $w \in L^F_{i + 1}$ for which $\ell(v)_p = \ell(w)_p$ for each $p \in \fK$.
    In fact, the $b$-balancedness of $T$ implies that there is precisely one way (up to isomorphism) to assign labels such that this property is satisfied.
    We define the label assignment of $F(T)$ to be this assignment.
\end{definition}

We will prove in \Cref{lem:iftthenft} that the tree $F(T)$ is indeed a construction tree.
However, before doing so, we will state a useful observation that in particular presents in a slightly more digestible way than \Cref{def:fcdot} how to infer the labels of nodes in some layer of $F(T)$ from the ``corresponding'' node in $T$.
The observation follows directly from \Cref{def:fcdot} (and in particular from the fact that $Q$ is obtained from $T$ in a depth-first search manner).
Recall the definition of the early rank of a node from \Cref{def:fcdot}.

\begin{observation}\label{obs:rephrasedigestibly}
    Let $v$ be a reflect node of $T$ with reflect index $i$ (which, by Property~\ref{prop:reflectlabellength} of \Cref{def:solidity}, in particular implies that $\len(v) = i$).
    Let $\fJ$ be the set of indices $j$ such that there is an ancestor $w \neq v$ of $v$ in $T$ that is a reflect node and has early rank $j$.
    Let $j_1 < j_2 < \dots < j_{i-1}$ denote the elements of $\fJ$.
    Let $a$ be the early rank of $v$ (which, by the definition of $Q$ in \Cref{def:fcdot}, in particular implies that $j < a$, for each $j \in \fJ$).
    Let $h$ be the early rank of the node with greatest early rank in the maximal subtree of $T$ hanging from $v$. 
    Let $v_p$ denote the first occurrence of $v$ in $Q$ and $v_q$ the second (which in particular implies that $v_p$ and $v_q$ are twins and that $p < q$).
    Let $\fK_p \subseteq \{1, \dots, |V_R|\}$, resp.\ $\fK_q \subseteq \{1, \dots, |V_R|\}$, be the set of indices $k$ such that $\fS^F_k$ has layer index $< p$, resp.\ $< q$, in $F(T)$.
    Then, the following hold.
    \begin{enumerate}
        \item The sets $\fJ$, $\fK_p$, and $\{ a \}$ are pairwise disjoint, and their union equals $\{ 1, \dots, a \}$.
        \item The sets $\fJ$, $\fK_q$, and $\{ a \}$ are pairwise disjoint, and their union equals $\{ 1, \dots, h \}$. 
        \item The $\fK_p$-padded version of $\ell(v) = \ell(v_p)$ is the string $y$ of length $a$ satisfying $y_0 = \ell(v)_0$, $y_{j_d} = \ell(v)_d$, for each $1 \leq d \leq i - 1$, and $y_k = \square$, for each $k \in \fK_p$.
        \item The $\fK_q$-padded version of $\ell(v) = \ell(v_q)$ is the string $y$ of length $h + 1$ satisfying $y_0 = \ell(v)_0$, $y_{j_d} = \ell(v)_d$, for each $1 \leq d \leq i - 1$, $y_a = *$, and $y_k = \square$, for each $k \in \fK_q$.
        \item The labels of the nodes in reflect layer $L^f_p = \fR^F_a$ of $F(T)$ are precisely those strings $y$ of length $a$ that satisfy $y_0 = \ell(v)_0$, $y_{j_d} = \ell(v)_d$, for each $1 \leq d \leq i - 1$, and $y_k \in \{ 1, \dots, b\}$, for each $k \in \fK_p$.
        \item The labels of the nodes in split layer $L^f_q = \fS^F_a$ of $F(T)$ are precisely those strings $y$ of length $h + 1$ that satisfy $y_0 = \ell(v)_0$, $y_{j_d} = \ell(v)_d$, for each $1 \leq d \leq i - 1$, $y_a = *$, and $y_k \in \{ 1, \dots, b\}$, for each $k \in \fK_q$.
    \end{enumerate}
\end{observation}

\begin{lemma}\label{lem:iftthenft}
    Let $T$ be a $b$-ary construction tree.
    Then $F(T)$ is a $b$-ary construction tree.
\end{lemma}
\begin{proof}
    As already observed in \Cref{def:fcdot}, $F(T)$ is a well-nested $b$-balanced rooted tree.
    It remains to show that it also satisfies the properties listed in \Cref{def:solidity}.

    Properties~\ref{prop:rootone}, \ref{prop:reflectlabellength}, \ref{prop:nosamereflect}, \ref{prop:reflectnostar}, and~\ref{prop:numberplusone} follow directly from the definition of $F(T)$.

    Next, observe that (the definition of $\fK$ and) the property used in \Cref{def:fcdot} for deciding for each layer which node is assigned which of the labels that have been reserved for that layer implies directly that $F(T)$ satisfies Property~\ref{prop:splitkids} of \Cref{def:solidity}.

    For proving Property~\ref{prop:splitstar}, consider some arbitrary split node $u \in V(F(T))$.
    Let $a$ be the split index of $u$, and $w$ the unique ancestor of $u$ in reflect layer $\fR^F_a$.
    Let $p$ be the layer index of $w$, and $q$ the layer index of $u$.
    From \Cref{def:fcdot}, we obtain that there must be some reflect node $v$ of $T$ such that the first occurrence of $v$ in $Q$ is $v_p$ and the second occurrence $v_q$.
    Now, by \Cref{obs:rephrasedigestibly} and Property~\ref{prop:splitkids} of \Cref{def:solidity} for $F(T)$, it follows that the $a$-th symbol of $\ell(u)$ is $*$ and the $0$-th to $(a-1)$-th symbols of $\ell(u)$ are the same as in $\ell(w)$.
    Moreover, by \Cref{def:fcdot}, Property~\ref{prop:reflectnostar} of \Cref{def:solidity} (for $T$), and the fact that $v$ is a reflect node, we obtain that $\ell(u)$ contains the symbol $*$ precisely once, which concludes the proof that $F(T)$ satisfies Property~\ref{prop:splitstar} of \Cref{def:solidity}.
    
    Now, we turn towards proving Property~\ref{prop:indepandclearing}, starting with the independence.
    Assume for a contradiction that the set of labels of all reflect nodes of $F(T)$ is not independent, and let $u$ and $w$ be two reflect nodes of $F(T)$ such that $\ell(u)$ is a final substring of $\ell(w)$.
    By Property~\ref{prop:nosamereflect} of \Cref{def:solidity} for $F(T)$, we know that $\ell(u) \neq \ell(w)$, which in particular implies that $u$ and $w$ are in different layers in $F(T)$ (as the labels of the nodes in any fixed layer have the same length).
    Let $i_u$ be the layer index of $u$ in $F(T)$ and $i_w$ the layer index of $w$.
    Recall the definition of $Q = (v_1, \dots, v_{2|V_R|})$ from \Cref{def:fcdot}.

    Let $u'$ be the reflect node in $T$ such that $v_{i_u}$ is the first occurrence of $u'$ in $Q$, and $w'$ the reflect node in $T$ such that $v_{i_w}$ is the first occurrence of $w'$ in $Q$ (which in particular implies that $u' \neq w'$).
    Let $v'$ denote the common ancestor of $u'$ and $w'$ in $T$ that has the largest distance from the root $r$.
    We consider three cases.

    First, consider the case that $u' \neq v' \neq w'$.
    Then $v'$ is a split node.
    Let $j$ be the split index of $v'$.
    By Property~\ref{prop:splitkids} of \Cref{def:solidity}, we obtain that $\len(\ell(u')) \geq j + 1$, $\len(\ell(w')) \geq j + 1$ and $\ell(u')_j \neq \ell(w')_j$.
    Moreover, by the well-nestedness of $T$, there is a (unique) ancestor of $v'$ in reflect layer $L_j$.
    Call this ancestor $\hat{v}$, and let $k$ denote the early rank of $\hat{v}$ in $Q$. 
    Then, by \Cref{obs:rephrasedigestibly} and the fact that $\hat{v}$ is an ancestor of both $u'$ and $w'$ in $T$ (with $u' \neq \hat{v} \neq w'$), we obtain $\ell(u)_k = \ell(u')_j$ and $\ell(w)_k = \ell(w')_j$.
    Since $\ell(u')_j \neq \ell(w')_j$ (as observed above), it follows that $\ell(u)_k \neq \ell(w)_k$, yielding a contradiction to the fact that $\ell(u)$ is a final substring of $\ell(w)$.

    Next, consider the case that $u' = v'$.
    Let $j$ be the reflect index of $u'$.
    Since $u' \neq w'$ and $\len(\ell(u')) = j$ (by Property~\ref{prop:reflectlabellength} of \Cref{def:solidity}), it follows, by Property~\ref{prop:indepandclearing} of \Cref{def:solidity}, that there is some $j' \leq j - 1$ such that $\ell(u')_{j'} \neq \ell(w')_{j'}$.
    Let $\hat{v}$ denote the unique ancestor of $u'$ in reflect layer $L_{j'}$, and let $k$ denote the early rank of $\hat{v}$ in $Q$.
    Then, analogously to the first case, we obtain $\ell(u)_k = \ell(u')_{j'} \neq \ell(w')_{j'} = \ell(w)_k$, yielding the desired contradiction.

    Finally, for the case that $w' = v'$, we obtain a contradiction analogously to the case $u' = v'$.
    This concludes the proof by contradiction and shows that the set of labels of all reflect nodes of $F(T)$ is independent.

    We continue by showing that this set is also clearing.
    Let $p$ denote the number of reflect nodes of $T$ and $q$ the number of reflect layers of $T$.
    Observe that, by the construction of $F(T)$, the number of reflect layers of $F(T)$ is also equal to $p$.
    By Property~\ref{prop:reflectlabellength} of \Cref{def:solidity} for $F(T)$, we obtain that the length of the set of labels of all reflect nodes of $F(T)$ (as defined in \Cref{def:indepclear}) is also equal to $p$.
    Consider an arbitrary string $z \in X_b^-$ of length $\geq p$.

    We define a sequence $W = (w_1, w_2, \dots)$ of nodes of $T$ and a sequence $Y = (y^{(1)}, y^{(2)}, \dots)$ of strings as follows.
    Set $w_1 := r$ and $y^{(1)} := z_0$ (which in particular implies that $w_1 \in \fR_1$, by \Cref{obs:basicconstree}).
    Then recursively do the following for $i = 1, 2, \dots, q - 1$.
    If $\ell(w_i) = y^{(i)}$, then terminate the recursion and set $w := w_i$ and $y := y^{(i)}$; otherwise do the following.
    Let $\fJ \subseteq \{ 1, \dots, q \}$ be the set of all indices $j$ such that the layer index of $\fS_j$ is strictly greater than the layer index of $\fR_i$ and strictly smaller than the layer index of $\fR_{i + 1}$.
    Set $y^{(i + 1)} := z_k y^{(i)}$ where $k$ is the early rank of $w_i$.
    Define $w_{i + 1}$ to be the unique descendant of $w_i$ in reflect layer $\fR_{i + 1}$ satisfying that the $j$-th symbol of $\ell(w_{i + 1})$ is equal to $y^{(i + 1)}_j$, for each $j \in \fJ$.
    (Such a descendant exists (and is unique) due to Property~\ref{prop:splitkids} of \Cref{def:solidity}.)
    If the recursion is not terminated before completing all recursion steps, set $w := w_q$ and $y := y^{(q)}$.

    Let $h$ be the early rank of $w$.
    We claim that there exists a node $u$ in reflect layer $\fR^F_h$ of $F(T)$ such that $\ell(u)$ is a final substring of $z$.
    
    Towards proving this claim, we first show that $\ell(w) = y$.
    If the above recursion terminates before completing all recursion steps, this follows directly from the definition of the recursive process.
    Hence, assume that the recursion does not terminate before completing all recursion steps, which implies that $w = w_q$ and $y = y^{(q)}$.
    By Property~\ref{prop:reflectlabellength} of \Cref{def:solidity}, it follows that $\len(y) = q = \len(\{ \ell(v) \mid v \in V_R \})$ (where $V_R$ denotes the set of reflect nodes of $T$).
    Hence, by Property~\ref{prop:indepandclearing} of \Cref{def:solidity}, there must be some reflect node $\hat{w}$ such that $\ell(\hat{w})$ is a final substring of $y$.
    
    If $\hat{w}$ does not lie on the path from $r$ to $w$, then there must be some split node that is an ancestor of both $w$ and $\hat{w}$ (since $w$ is contained in the final reflect layer $\fR_q$); let $w'$ denote the split node of this kind with largest distance from the root $r$.
    But then, for the split index $a$ of $w'$, the construction of $y$ (together with Property~\ref{prop:splitkids}) guarantees that $\ell(\hat{w})_a \neq y_a$, contradicting the fact that $\ell(\hat{w})$ is a final substring of $y$.
    Hence, assume that $\hat{w}$ lies on the path from $r$ to $w$, and let $\hat{i}$ be the reflect index of $\hat{w}$.
    By the definition of the recursive process, we obtain that $\hat{w} = w_{\hat{i}}$ and $\ell(\hat{w}) = y^{\hat{i}}$, which implies that the recursion terminates after the recursion step for $i = \hat{i} - 1$.
    As the recursion does not terminate before completing all recursion steps, we obtain that $\hat{i} = q$ and $\ell(w) = \ell(\hat{w}) = y^{\hat{i}} = y$. 

    Let $g$ denote the reflect index of $w$ and recall that $h$ denotes the early rank of $w$.
    From the definition of the recursive process above, it follows that the reflect nodes of $T$ that are ancestors of $w$ are precisely $r = w_1, w_2, \dots, w_g = w$, ordered by increasing distance from the root $r$.
    Let $h_d$ denote the early rank of $w_d$, for each $1 \leq d \leq g - 1$.    
    By \Cref{obs:rephrasedigestibly}, the fact that $\ell(w) = y$, and the definition of the $y^{(i)}$ in the recursive process, we obtain that the labels of the nodes in reflect layer $\fR^F_h$ of $F(T)$ are precisely those strings $x \in X_b^-$ of length $h$ that satisfy $x_0 = y_0 = z_0$ and $x_{h_d} = y_d = z_{h_d}$, for each $1 \leq d \leq g - 1$.
    Since $h \leq p$, it follows that there exists some node $u \in \fR^F_h$ such that $\ell(u)$ is a final substring of $z$.    
    This shows that the set of labels of all reflect nodes of $F(T)$ is clearing and concludes the proof that $F(T)$ satisfies Property~\ref{prop:indepandclearing} of \Cref{def:solidity}.
\end{proof}

\minusspace
\paragraph{The Construction Tree Sequence.}

Now we are set to finally define the desired sequence $T_2, T_3, \dots$ of construction trees (where the fact that we start with index $2$ is for technical reasons).
Please note that the construction tree from~\Cref{fig:constructiontree} is actually $T_2$ for $\Delta = 3$.

\begin{definition}\label{def:constreesequence}
We define $T_2$ as follows.
Tree $T_2$ has $\Delta^3 + 2\Delta^2 + 2\Delta + 3$ nodes, with labels
\begin{itemize}
    \item $1, 12, *12$,
    \item $j22, *j22$ for each $1 \leq j \leq \Delta$,
    \item $kj32, *kj32$ for each $1 \leq j,k \leq \Delta$, and
    \item $pkj$$*$$1$ for each $1 \leq j,k,p \leq \Delta$.
\end{itemize}
The node with label $1$ is the root of $T_2$, and the edges of $T_2$ are as follows, where for simplicity, we represent a node by its label.
\begin{itemize}
    \item $(1, 12), (12, *12)$,
    \item $(*12, j22), (j22, *j22)$, for each $1 \leq j \leq \Delta$,
    \item $(*j22, kj32), (kj32, *kj32)$ for each $1 \leq j,k \leq \Delta$, and
    \item $(*kj32,pkj$$*$$1)$ for each $1 \leq j,k,p \leq \Delta$.
\end{itemize}

Moreover, for any integer $i \geq 3$, we set $T_i := F(T_{i - 1})$.
\end{definition}

The reflect nodes of $T_2$ are those that do not contain the symbol $*$; the split nodes are those that do contain the symbol $*$.
It is straightforward to verify that $T_2$ is a well-nested $\Delta$-balanced rooted tree with a solid labeling, i.e., that $T_2$ is a $\Delta$-ary construction tree.
By~\Cref{lem:iftthenft}, this implies that $T_i$ is a $\Delta$-ary construction tree, for any integer $i \geq 2$.

\minusspace
\paragraph{Bounding the number of nodes.}
Next we bound the number of nodes of the trees $T_i$.
To this end, we introduce the following notation, capturing certain power towers in a convenient format.

\begin{definition}\label{def:powertower}
	Let $j$, $k$, and $\Delta \geq 3$ be positive integers.
	We denote the power tower of height\footnote{We say that a power tower has height $j$ if it consists of $j$ entries; e.g., the power tower $5^{6^7}$ has height $3$.} $j$ with the top exponent being $k$ and the other $j - 1$ entries being $\Delta$ by $P_{\Delta}(j,k)$.
	(For instance, $P_{\Delta}(3, 5) = \Delta^{\Delta^5}$.)
\end{definition}

Now, for each positive integer $i$, we upper bound the number of nodes of $T_i$.
\begin{lemma}\label{lem:numbernodesub}
    For each integer $i \geq 2$ (and any $\Delta \geq 3$), the number of nodes of $T_i$ is at most
    $P_{\Delta}(i, \Delta + 1)$.
\end{lemma}
\begin{proof}
	From the definition of the $T_i$ (and the fact that, due to its well-nestedness, any construction tree contains at least as many split nodes as reflect nodes), we obtain directly that, if $k$ denotes the number of nodes of tree $T_i$, then $T_{i + 1}$ has at most $\Delta^{k/2}$ leaf nodes.
	This in turn implies that $T_{i + 1}$ has at most $2\Delta^{k/2}$ split nodes, which implies that the total number of nodes of $T_{i + 1}$ is at most $4\Delta^{k/2}$, due to the well-nestedness of $T_{i + 1}$.
	Note that, for any $k \geq 4$ (and any $\Delta \geq 3$), we have $\Delta^{k/2} > 4$, which implies $\Delta^{k} > 4\Delta^{k/2}$.
	Together with the fact that the number of nodes of $T_2$ is $\Delta^3 + 2\Delta^2 + 2\Delta + 3$ (and the fact that $\Delta^3 + 2\Delta^2 + 2\Delta + 3 \leq \Delta^4 \leq \Delta^{\Delta + 1}$ for any $\Delta \geq 3$), the above discussion implies the lemma statement.
\end{proof}

In the following, we show that the number of nodes of the $T_i$ is not far away from the upper bound presented in~\Cref{lem:numbernodesub}.
Please note that this is not necessary for obtaining the lower bound we present in this work; the purpose of the following lemma is to imply that we cannot obtain a stronger lower bound (asymptotically in the exponent) by improving the analysis in the proof of~\Cref{lem:numbernodesub}. 

\begin{lemma}\label{numbernodeslb}
    For each integer $i \geq 2$ (and any $\Delta \geq 3$), the number of nodes of $T_i$ is at least $P_{\Delta}(\lfloor i/2 \rfloor, \Delta + 1)$.
\end{lemma}
\begin{proof}
	We start by observing that the definition of $F(\cdot)$ ensures that if $k$ denotes the number of nodes in the last reflect layer of some construction tree $T_i$, then the number of nodes in the last reflect layer of $T_{i + 1}$ is at least $\Delta^{k - 1}$ (as, in $T_{i + 1}$, there must be at least $k - 1$ split layers with smaller layer index than the last reflect layer).
	Since $\Delta^{k - 1} - 1 \geq k$ (for any $k \geq 2$ and $\Delta \geq 3$), we obtain that the number of nodes in the last reflect layer of $T_{i + 2}$ is at least $\Delta^k$.
	As the number of reflect nodes in the last reflect layer of $T_2$ is $\Delta^2 \geq \Delta + 1$, the lemma statement follows.
\end{proof}

\subsection{Auxiliary Labelings of Construction Trees}\label{sec:aux}
Before describing (in~\Cref{sec:inputtree}) how the information stored in a construction tree is used to create an actual input instance, we introduce two auxiliary labelings of construction trees that will be highly useful for proving statements about the input instance and related objects.
More specifically, we introduce two auxiliary labelings of the edges of the construction tree $T$ and collect some basic properties of the auxiliary labelings.
On a high level, these two labelings capture information about the structure and labels of a certain sequence of objects that we will use to derive the input tree from the construction tree in~\Cref{sec:inputtree}.

We start by introducing the first of the two labelings, which we call $\psi$.
Recall the definition of $X_b^-$ from \Cref{def:indepclear}.

\begin{definition}[edge labeling $\psi$]\label{def:edgelabeling}
    Let $T$ be a $\Delta$-ary construction tree.
    We define an edge labeling $\psi: E(T) \rightarrow 2^{X_{\Delta}^-}$ that assigns to each edge of $T$ a finite (and possibly empty) subset of $X_{\Delta}^-$ via the following inductive process.
    
    Let $e$ be the edge whose head is the root $r$ of $T$.
    (Note that \Cref{obs:basicconstree} asserts that $r$ is a reflect node, which implies that $r$ has only one incoming edge.)
    Then we set $\psi(e) := \{ i2 \mid 1 \leq i \leq \Delta \}$.

    Now, let $e = (u, v)$ be any edge of $T$ not connected to the root and assume that the unique edge $e' = (v, w)$ with tail $v$ has already been labeled w.r.t.\ $\psi$.
    If $v$ is a reflect node, then we set $\psi(e) := \{ iz \mid 1 \leq i \leq \Delta, z \in \psi(e'), z \neq \ell(v) \}$.
    If $v$ is a split node, then we set $\psi(e) := \{ z \mid z \in \psi(e'), z_j = i \}$ where $j$ is the index such that the $j$-th symbol of $\ell(v)$ is $*$ (which is unique by Property~\ref{prop:splitstar} of \Cref{def:solidity}), and $i \in \{ 1, \dots, \Delta \}$ is the $j$-th symbol of $\ell(u)$ (i.e., $(\ell(v))_j = *$ and $\ell(u)_j = i$).
    (Note that it is impossible that $u_j = *$ due to Properties~\ref{prop:splitstar} and~\ref{prop:splitkids} of \Cref{def:solidity}.)
\end{definition}

Next, we prove that the node labeling of $T$ is related to the edge labeling $\psi$ via a useful property.

\begin{lemma}\label{lem:nodeandedge}
    Let $v \neq r$ be a reflect node in $T$ and $e = (v, w)$ the edge leading to its parent $w$.
    Then $\ell(v) \in \psi(e)$.
\end{lemma}
\begin{proof}
    Set $z := \ell(v)$.
    Consider the path $r = u_1, u_2, \dots, u_k = v$ connecting $r$ and $v$.
    For each $2 \leq i \leq k$, denote the number of reflect nodes in $\{ u_1, \dots, u_{i - 1} \}$ by $\Lambda_i$.
    Observe that the length of $z$ is $\Lambda_k + 1$, by Property~\ref{prop:reflectlabellength} of \Cref{def:solidity}.
    Furthermore, for each $2 \leq i \leq \Lambda_k + 1$, set $z^{(i)} := z_{i - 1}z_{i - 2} \dots z_0$ to be the final substring of $z$ of length $i$.
    We prove by induction that, for each $2 \leq i \leq k$, the string $z^{(\Lambda_i + 1)}$ is contained in $\psi((u_i, u_{i - 1}))$.
    
    For the base case $i = 2$, it suffices to observe that $z^{(\Lambda_i + 1)} = z^{(2)}$ is contained in $\psi((u_2, u_1)) = \{ 12, 22, \dots, \Delta2 \}$ since the $0$-th symbol of $z$ is $2$ (where we make use of \Cref{obs:basicconstree}).
    For the induction step, assume that the induction hypothesis holds for $i = p$ (where $2 \leq p \leq k - 1$).
    We show that it then also holds for $i = p + 1$.

    Consider first the case that $u_p$ is a reflect node.
    Then $\Lambda_{p + 1} = \Lambda_p + 1$, and the string $z^{(\Lambda_{p + 1} + 1)} = z^{(\Lambda_{p} + 2)}$ is contained in the set $\{ jz^{(\Lambda_{p} + 1)} \mid 1 \leq j \leq \Delta\}$.
    By the induction hypothesis for $i = p$, we know that $z^{(\Lambda_p + 1)}$ is contained in $\psi((u_p, u_{p - 1}))$.
    Moreover, by Property~\ref{prop:indepandclearing} of \Cref{def:solidity}, we know that $\ell(u_p)$ is not a final substring of $z$, which implies $z^{(\Lambda_p + 1)} \neq \ell(u_p)$.
    It follows that $z^{(\Lambda_{p + 1} + 1)}$ is contained in $\{ jy \mid 1 \leq j \leq \Delta, y \in \psi((u_p, u_{p - 1})), y \neq \ell(u_p) \} = \psi((u_{p + 1}, u_p))$, as desired.

    Now consider the case that $u_p$ is a split node.
    Since $p < k$, node $u_p$ is an internal split node.
    Let $j$ be the index such that the $j$-th symbol of $\ell(u_p)$ is $*$ (which is unique by Property~\ref{prop:splitstar} of \Cref{def:solidity}).
    By Property~\ref{prop:splitkids} of \Cref{def:solidity}, we know that the $j$-th symbol of $\ell(u_{p + 1})$ is identical to $z_j$, which, by the definition of $\psi$, implies that $\psi((u_{p + 1}, u_p)) = \{ y \mid y \in \psi((u_p, u_{p - 1})), y_j = z_j \}$.
    Since $\Lambda_{p + 1} = \Lambda_p$, it follows by the induction hypothesis that $z^{(\Lambda_{p + 1} + 1)} = z^{(\Lambda_p + 1)}$ is contained in $\psi((u_{p + 1}, u_p))$, as desired.
    This concludes the proof by induction.

    By setting $i := k$, we obtain that $\ell(v) = z^{(\Lambda_k + 1)}$ is contained in $\psi(e) = \psi((u_k, u_{k - 1}))$.
\end{proof}

Now, we define the second of the mentioned edge labelings.

\begin{definition}[edge labeling $\pi$]\label{def:edgelabelingtwo}
    Let $T$ be a $\Delta$-ary construction tree.
    We define an edge labeling $\pi: E(T) \rightarrow 2^{X_{\Delta}}$ that assigns to each edge of $T$ a finite subset of $X_{\Delta}$ via the following inductive process.
    
    Let $e$ be the edge whose head is the root $r$ of $T$.
    Then we set $\pi(e) := \{ *1 \}$.

    Now, let $e = (u, v)$ be any edge of $T$ not connected to the root and assume that the unique edge $e' = (v, w)$ with tail $v$ has already been labeled w.r.t.\ $\pi$.
    If $v$ is a reflect node, then we set $\pi(e) := \{ iz \mid 1 \leq i \leq \Delta, z \in \pi(e'), z \neq \ell(v) \} \cup \{ *\ell(v) \}$.\footnote{In fact, one can show that $z \in \pi(e')$ implies $z \neq \ell(v)$ but at this point it is easier to explicitly add the condition $z \neq \ell(v)$.}
    If $v$ is a split node, then we set $\pi(e) := \{ z \mid z \in \pi(e'), z_j = i \}$ where $j$ is the index such that the $j$-th symbol of $\ell(v)$ is $*$ (which is unique by Property~\ref{prop:splitstar} of \Cref{def:solidity}), and $i \in \{ 1, \dots, \Delta \}$ is the $j$-th symbol of $\ell(u)$ (i.e., $(\ell(v))_j = *$ and $\ell(u)_j = i$).
\end{definition}

Analogous to the statement~\Cref{lem:nodeandedge} makes about $\psi$, the following lemma shows a relation between $\pi$ and the node labeling of $T$.

\begin{lemma}\label{lem:nodeandedgetwo}
    Let $v \neq r$ be a split node in $T$ and $e = (v, w)$ the edge leading to its parent $w$. Then $\ell(v) \in \pi(e)$.
\end{lemma}
\begin{proof}
    Let $i$ be the index such that $v \in \fS_i$ and let $u$ be the unique predecessor of $v$ in reflect layer $\fR_i$.
    Set $z := \ell(v)$ and $y := \ell(u)$.
    By Property~\ref{prop:splitstar} of \Cref{def:solidity}, we know that $z_i = *$ and $z_j = y_j$, for each $0 \leq j \leq i - 1$.
    Consider the path $u = u_1, u_2, \dots, u_k = v$ connecting $u$ and $v$.

    Similar to the proof of \Cref{lem:nodeandedge}, for each $1 \leq p \leq k$, denote the number of reflect nodes on the path connecting the root $r$ to the parent of $u_p$ by $\Lambda_p$.
    Observe that the length of $z$ is $\Lambda_k + 1$, by Property~\ref{prop:numberplusone} of \Cref{def:solidity}, and that the length of $y$ is $\Lambda_1 + 1 = i$, by Property~\ref{prop:reflectlabellength} of \Cref{def:solidity}, which in particular implies that $y$ is a final substring of $z$.
    Furthermore, for each $2 \leq p \leq \Lambda_k + 1$, set $z^{(p)} := z_{p - 1}z_{p - 2} \dots z_0$ to be the final substring of $z$ of length $p$.
    We prove by induction that, for each $2 \leq p \leq k$, the string $z^{(\Lambda_p + 1)}$ is contained in $\pi((u_p, u_{p - 1}))$.

    For the base case $p = 2$, it suffices to observe that the definition of $\pi$ together with $\len(y) = \Lambda_1 + 1 = \Lambda_2$ implies that $z^{(\Lambda_2 + 1)} = *y$ is contained in $\pi((u_2, u_{1}))$.
    For the induction step, assume that the induction hypothesis holds for $p = q$ (where $2 \leq q \leq k - 1$).
    We show that it then also holds for $p = q + 1$.

    Consider first the case that $u_q$ is a reflect node.
    By Property~\ref{prop:indepandclearing} of \Cref{def:solidity}, we know that $\ell(u_q)$ is not a final substring of $y$ and vice versa, which, by the fact that $y$ is a final substring of $z$, implies that $\ell(u_q)$ is not a final substring of $z$ (and vice versa).
    Hence, the fact that $\Lambda_{q + 1} = \Lambda_q + 1$ and the induction hypothesis for $p = q$, together with the definition of $\pi$, directly imply that $z^{(\Lambda_{q + 1} + 1)} = z^{(\Lambda_q + 2)}$ is contained in $\pi((u_{q + 1}, u_{q}))$.

    Now consider the case that $u_q$ is a split node.
    Since $q < k$, node $u_q$ is an internal split node.
    Let $j$ be the index such that the $j$-th symbol of $\ell(u_q)$ is $*$ (which is unique by Property~\ref{prop:splitstar} of \Cref{def:solidity}).
    By Property~\ref{prop:splitkids} of \Cref{def:solidity}, we know that the $j$-th symbol of $\ell(u_{q + 1})$ is identical to $z_j$, which, by the definition of $\pi$, the induction hypothesis, and the fact that $\Lambda_{q + 1} = \Lambda_{q}$ implies that $z^{(\Lambda_{q + 1} + 1)} = z^{(\Lambda_q + 1)}$ is contained in $\pi((u_{q + 1}, u_{q}))$.
    This concludes the proof by induction.

    By setting $p := k$, we obtain that $\ell(v) = z^{(k + 1)}$ is contained in $\pi(e) = \pi((u_k, u_{k - 1}))$.
\end{proof}

Next, we examine the edge labels on the edges incident to leaf nodes of $T$.

\begin{lemma}\label{lem:psiandpileaf}
    Let $v$ be a leaf node of $T$ and $e$ the unique edge incident to $v$.
    Then $\psi(e) = \emptyset$ and $\pi(e) = \{ \ell(v) \}$.
\end{lemma}
\begin{proof}
    We start by showing $\psi(e) = \emptyset$.
    For a contradiction assume that $\psi(e) \neq \emptyset$, and let $z \in X_{\Delta}^-$ be some string contained in $\psi(e)$.
    Let $k$ denote the number of reflect layers in $T$, which, by the definition of $\psi$ (and \Cref{obs:basicconstree}) implies that $\len(z) = k + 1$ and that the label of any reflect node of $T$ has length at most $k$.
    By Property~\ref{prop:indepandclearing} of \Cref{def:solidity}, it follows that there is some reflect node $u$ of $T$ such that $\ell(u)$ is a final substring of $z$.
    Let $w$ denote the common ancestor of both $u$ and $v$ that has the largest distance from the root $r$, which in particular implies that $w$ is a split node or $u = w$.
    
    Consider first the case that $u \neq w$, and let $i$ be the index such that the $i$-th symbol of $\ell(w)$ is $*$.
    Then the definition of $\psi$, together with Properties~\ref{prop:splitstar} and~\ref{prop:splitkids} of \Cref{def:solidity}, implies that the $i$-th symbol of $\ell(u)$ and the $i$-th symbol of $z$ (exist and) differ, violating the fact that $\ell(u)$ is a final substring of $z$.

    Now consider the case that $u = w$, which implies that $u$ is an ancestor of $v$.
    Observe that the design of $\psi$ ensures that $\psi(e)$ contains only strings whose $0$-th symbol is $2$, which implies that $u \neq r$ (by the definition of $u$).
    Let $e'$ be the edge with tail $u$, and $e''$ the edge with head $u$.
    By the definition of $\psi$ and Property~\ref{prop:reflectlabellength} of \Cref{def:solidity}, the length of all strings in $\psi(e')$ is equal and identical to the length of $\ell(u)$.
    This implies, again due to the definition of $\psi$, that $\psi(e'')$ does not contain any string of which $\ell(u)$ is a final substring.
    Now observe that the definition of $\psi$ guarantees that if, for some edge $f$ and some string $y$, it holds that there is no substring in $\psi(f)$ of which $y$ is a final substring, then this also holds for any edge $f'$ whose head is the tail of $f$ (for the same $y$).
    By applying this argument inductively along the path from $u$ to $v$, the fact that $u$ is an ancestor of $v$ implies that $\ell(u)$ is not a final substring of $z$, yielding a contradiction and concluding the proof that $\psi(e) = \emptyset$.

    Next, we show that $\pi(e) = \{ \ell(v) \}$.
    For a contradiction, assume that $\pi(e) \neq \{ \ell(v) \}$.
    Since $\ell(v) \in \pi(e)$ (by \Cref{lem:nodeandedgetwo}), there must exist some string $z$ such that $z \in \pi(e)$ and $z \neq \ell(v)$.
    As before, let $k$ denote the number of reflect layers in $T$.
    Then the design of $\pi$ (together with \Cref{obs:basicconstree}) ensures that $\len(z) = k + 1$ and that there exists some index $1 \leq i \leq k$ such that the $i$-th symbol of $z$ is $*$.
    Let $u$ be the unique ancestor of $v$ in split layer $\fS_i$ and $w$ the unique ancestor of $v$ in reflect layer $\fR_i$ (which also implies that $w$ is an ancestor of $u$).
    Let $e'$ denote the edge with head $w$ and $e''$ the edge with tail $u$.
    By \Cref{def:solidity} and the definition of $\pi$, there is precisely one string $y \in \pi(e')$ satisfying that the $i$-th symbol of $y$ is $*$.
    Moreover, by the well-nestedness of $T$ (see~\Cref{def:wellnested}), \Cref{def:solidity}, and the definition of $\pi$, we obtain that the only string in $\pi(e'')$ whose $i$-th symbol can possibly be $*$ is $\ell(u)$.
    This implies that, for each edge $e'''$ with head $u$, the set $\pi(e''')$ does not contain a string whose $i$-th symbol is $*$ (again, by the definition of $\pi$).
    Now, the definition of $\pi$ guarantees (similarly to our considerations for $\psi$) that this is also true for any edge whose tail is a descendant of $u$.
    In particular, we obtain that $\pi(e)$ does not contain $z$, yielding a contradiction, and concluding the proof.   
\end{proof}

Lastly, we prove a technical lemma that relates the labels on an edge whose tail is some split node $v$ to the labels on the edge whose head is the reflect node ``corresponding'' to $v$.

\begin{lemma}\label{lem:compbeforesplit}
    Let $v$ be a split node of $T$ with split index $i$ and let $u$ be the unique ancestor of $v$ in reflect layer $\fR_i$.
    Let $e$ denote the edge with head $u$ and $e'$ the edge with tail $v$.
    Let $w_1, \dots, w_k$ denote the split nodes on the path from $u$ to $v$ (excluding $v$, and ordered in increasing distance from the root).
    For each $1 \leq j \leq k$, let $p_j$ denote the split index of $w_j$ and $q_j$ the element from $\{ 1, \dots, \Delta \}$ satisfying that the child of $w_j$ that lies on the path from $u$ to $v$ is the $q_j$-th child of $w_j$.
    Set $Y := \{ y_{i + k} \dots y_0 \mid y_{p_j} = q_j, \text{ for each $1 \leq j \leq k$, and } y_{i} \dots y_0 \in \psi(e) \cup \pi(e) \}$.\footnote{Note that $Y$ is well-defined, by the well-nestedness of $T$, the definitions of $\psi$ and $\pi$, and Property~\ref{prop:reflectlabellength} of \Cref{def:solidity}.}
    Then $\psi(e') \cup \pi(e') \subseteq Y$. 
\end{lemma}
\begin{proof}
    By the definitions of $\psi$ and $\pi$ and \Cref{lem:nodeandedgetwo}, for any two edges $f$ and $f'$ such that the tail of $f$ is the head of $f'$, it holds that, for any string $z' \in \psi(f') \cup \pi(f')$, there exists some string $z \in \psi(f) \cup \pi(f)$ such that $z$ is a final substring of $z'$.
    Applying this argument inductively along the path from $u$ to $v$, we obtain that for each string $x' \in \psi(e') \cup \pi(e')$, there must be some string $x \in \psi(e) \cup \pi(e)$ such that $x$ is a final substring of $x'$.
    Since each string in $\psi(e) \cup \pi(e)$ has length $i + 1$ (by the definitions of $\psi$ and $\pi$, and Property~\ref{prop:reflectlabellength} of \Cref{def:solidity}), we obtain that for each string $x' \in \psi(e') \cup \pi(e')$, we have $x'_{i} \dots x'_0 \in \psi(e) \cup \pi(e)$.
    Moreover, by the well-nestedness of $T$ and the definitions of $\psi$ and $\pi$, we also obtain for each string $x' \in \psi(e') \cup \pi(e')$ that the length of $x'$ is $i + k + 1$, and that $x'_{p_j} = q_j$, for each $1 \leq j \leq k$.
    It follows that $\psi(e') \cup \pi(e') \subseteq Y$.
\end{proof}

\subsection{The Input Tree}\label{sec:inputtree}
In this section, we show how to use the information stored in a construction tree to derive an input tree from which the actual lower bound instances we use in~\Cref{sec:lowerbound} can be obtained by minor adaptations (and by combining it with a suitable sequence of queries in the randomized online-LOCAL model).

Let $T$ be a construction tree.
We start by showing how to construct from $T$ a new tree $G_T$ (with node labels) that will be the aforementioned input tree.
To construct $G_T$ from $T$, we will make use of two operations---reflecting and splitting.
Before defining these two operations, we define the notion of a \emph{marked tree}.

\begin{definition}[marked tree]
    Let $\Delta \geq 3$ be some integer.
    A \emph{marked tree} is a triple $(G, \ell, W)$ where $G$ is a tree of maximum degree $\Delta$, $\ell: V(G) \rightarrow X_{\Delta}$ is a labeling of the nodes of $G$ with labels from $X_{\Delta}$, and $W \subseteq V(G)$ is a subset of the nodes of $G$.
    We call the nodes in $W$ \emph{marked} and the nodes in $V(G) \setminus W$ unmarked.
    For simplicity, we may omit $\ell$ and $W$ and use $G$ instead of $(G, \ell, W)$ when considering a marked graph.

    Moreover, we call a leaf (resp.\ node) $v$ of $G$ a \emph{$2$-leaf (resp.\ $2$-node)} if the $0$-th symbol of $\ell(v)$ is $2$, and a \emph{$1$-leaf (resp.\ $1$-node)} if the $0$-th symbol of $\ell(v)$ is $1$.
    Finally, we call nodes of $G$ that are not leaves \emph{internal nodes}.
\end{definition}

Now we are set to define the reflection and split operations. 

\begin{definition}[reflection]\label{def:reflect}
    Let $G$ be a marked tree and $v$ an unmarked $2$-leaf of $G$.
    Consider the maximal connected components of the subgraph of $G$ induced by the set of all unmarked nodes.
    Let $H_1$ denote the maximal connected component that contains $v$.
    In the following, we define a new tree, called $R_v(G)$, that we say is \emph{obtained from $G$ by reflection at $v$}.
    
    Start by creating $\Delta - 1$ copies $H_2, \dots, H_{\Delta}$ of $H_1$ (which includes copying node labels and whether a node is marked or unmarked).
    Next, for each $1 \leq i \leq \Delta$ and each $w \in V(H_i)$ that is neither $v$ nor a copy of $v$, replace the label $\ell(w)$ of $w$ by the string obtained by prepending symbol $i$ to $\ell(w)$.
    After this, ``join'' the graphs $G, H_2, \dots, H_{\Delta}$ by identifying $v \in V(H_1) \subseteq V(G)$ with its copies in the other $H_i$.
    (Note that all identified nodes had the same label and were all unmarked, which we will assume to be preserved during the identification.)
    Next, replace the label $\ell(v)$ of $v$ by the string obtained by prepending symbol $*$ to $\ell(v)$.
    Finally, change the port numbers at $v$ so that, for each $1 \leq i \leq \Delta$, the port number that $v$ assigns to its incident edge in $H_i$ is $i$.
    The obtained tree is the aforementioned tree $R_v(G)$.
    
    When convenient, we will consider $R_v(G)$ as a supergraph of $G$ with partially changed labels in the natural way.
    Moreover, for each node $w \neq v$ in $H_1$, we call its copy in $H_i$ the \emph{$i$-copy of $w$}, for each $1 \leq i \leq \Delta$.
\end{definition}

\begin{definition}[split]\label{def:split}
    Let $G$ be a marked tree and $v$ an unmarked node of $G$.
    We define the tree $S_v(G)$ as the tree obtained from $G$ by marking $v$ and say that $S_v(G)$ is \emph{obtained from $G$ by a split at $v$}.
    
    When convenient, we will consider $S_v(G)$ as a supergraph of $G$ with a slightly changed set of marked nodes in the natural way.
\end{definition}

Note that it directly follows from the definitions of the reflection and split operations that $R_v(G)$ and $S_v(G)$ are marked trees.
Given these operations, we are now ready to define the tree $G_T$.

\begin{definition}[$G_T$]\label{def:geetee}
    Let $\Delta \geq 3$ be an integer and let $T$ be a $\Delta$-ary construction tree.
    Let $\vo, \vt, \dots, v^{(|V(T)|)}$ be an arbitrary ordering of the nodes in $V(T)$ such that, for any $1 \leq i < j \leq |V(T)|$, either $\vj$ is a descendant of $\vi$ or $\vj$ and $\vi$ are incomparable.
    In other words, a node appears in the sequence only after all of its ancestors have appeared.
    Now, the marked tree $G_T$ is the tree obtained in the following way.

    Start with the marked tree $\Gz$ consisting of two unmarked nodes $v, w$ connected by an edge with labels $\ell(v) = 1$ and $\ell(w) = 2$.
    Then, modify $\Gz$ iteratively by following the ``instructions'' written in the construction tree $T$.
    More precisely, construct a sequence $\Gz, \Go, \dots, G^{(|V(T)|)}$ as follows.

    For each $1 \leq i \leq |V(T)|$, we define $\Gi := R_{\vi}(G^{(i-1)})$ if $\vi$ is a reflect node and $\Gi := S_{\vi}(G^{(i-1)})$ if $\vi$ is a split node.
    In other words, starting with $\Gz$, we process the nodes in the aforementioned node sequence one by one, each time performing a reflect or split operation at the processed node in the currently obtained marked tree, where the choice whether a reflect or split operation is performed is given by the type of the processed node in the construction tree.

    Finally, set $G_T := G^{(|V(T)|)}$.
    We call $G_T$ the \emph{input tree induced by $T$}.
\end{definition}

From the above description, it is not obvious that the tree $G_T$ is well-defined: to this end, we have to show that for each indicated reflection and split operation, the node at which the reflection/split is supposed to happen actually exists and is unique\footnote{Here, ``existence and uniqueness'' is meant in the sense that there is precisely one node in the currently obtained graph whose label is identical to the label that the node at which the reflection/split is supposed to happen has in the construction tree $T$.} in the currently obtained graph (and is an unmarked $2$-leaf in the case of a reflection operation and unmarked in the case of a split operation), and that the resulting graph $G_T$ is independent of the choice of the sequence $\vo, \vt, \dots, v^{(|V(T)|)}$.
We will take care of this in the following lemma.
(Note that the fact that the nodes at which an operation is to be performed actually exist in the respectively obtained graphs and have the required properties will then also imply that the graphs $\Gz, \Go, \dots, G^{(|V(T)|)}$ are all marked trees.)

\begin{lemma}\label{lem:welldef}
    The tree $G_T$ is well-defined.
    In particular, $G_T$ is independent of the choice of $\vo, \vt, \dots, v^{(|V(T)|)}$.
\end{lemma}
\begin{proof}
    We start by showing that whenever the described construction of $G_T$ specifies performing a reflection or a split, the node at which the reflection/split is supposed to happen actually exists and is unique in the obtained graph and that the node is an unmarked $2$-leaf in the case of a reflection operation and unmarked in the case of a split operation.
    Our proof proceeds by induction.
    More precisely, for any fixed sequence $\vo, \vt, \dots, v^{(|V(T)|)}$, we will prove the following statement by induction on $i$.
    
    For each $1 \leq i \leq |V(T)|$, the following hold:
    \begin{enumerate}
        \item\label{one} $G^{(i - 1)}$ contains precisely one node with label $\ell(\vi)$.
        \item\label{two} If $\vi$ is a reflect node, then $\vi$ (i.e., the unique node with label $\ell(\vi)$ in $G^{(i - 1)}$) is an unmarked leaf in $G^{(i - 1)}$ that is a $1$-leaf if $i = 1$ and a $2$-leaf if $i > 1$; if $\vi$ is a split node, then $\vi$ is unmarked in $G^{(i - 1)}$.
        \item\label{three} No two nodes of $\Gi$ have the same label and the set of labels of the nodes of $\Gi$ is independent (as defined in \Cref{def:indepclear}).
        \item\label{four} Let $\fC^{(i)}$ denote the set of maximal connected components in the subgraph of $\Gi$ induced by the unmarked nodes.
        Let $\fE^{(i)}$ denote the set of edges of $T$ with one endpoint in $\{ \vo, \dots, \vi \}$ and one endpoint in $\{ v^{(i + 1)}, \dots, v^{(|V(T)|)}\}$.
        Then there is a bijection $\alpha_i: \fC^{(i)} \rightarrow \fE^{(i)}$ such that, for each component $C \in \fC^{(i)}$,
        \begin{enumerate}
            \item\label{fouraa} the set of labels of the nodes of $C$ that are $2$-leaves of $G^{(i)}$ is precisely $\psi(\alpha_i(C))$, and
            \item\label{fourbb} the set of labels of the nodes of $C$ that are not $2$-leaves of $G^{(i)}$ is precisely $\pi(\alpha_i(C))$.
        \end{enumerate}
        \item\label{five} Let $\fC^{(i)}$ be defined as in Property~\ref{four}. Then, for each component $C \in \fC^{(i)}$ and each $2$-node $u$ of $C$, either $u$ is a leaf of $\Gi$ or $u$ has degree $\Delta$ in $C$.
        Moreover, all neighbors of each $2$-node in $\Gi$ are $1$-nodes and vice versa. 
    \end{enumerate}

    For the base case, consider $i = 1$.
    As $\Gz$ contains precisely one node with label $1$ and for the node $v^{(1)} = r$ in $T$ we have $\ell(\vo) = \ell(r) = 1$, Property~\ref{one} is satisfied.
    Since the node with label $1$ in $T$ is a reflect node and the node with label $1$ is an unmarked $1$-leaf in $\Gz$, also Property~\ref{two} is satisfied.
    Moreover, it is straightforward to verify that the tree $\Go$ obtained from $\Gz$ by reflection at the node with label $1$ contains only unmarked nodes and has $\Delta$ leaves labeled $12, 22, \dots, \Delta2$, as well as one internal node with label $*1$, which implies Properties~\ref{three} and~\ref{five}.
    Similarly, it is straightforward to verify that the unique edge in $T$ with head $r$ is labeled with the set $\{ 12, 22, \dots, \Delta2 \}$ under $\psi$ and with the set $\{ *1 \}$ under $\pi$, which, together with the above, implies Property~\ref{four}.

    For the induction step, assume that the induction hypothesis holds for each $i \leq k$ (where $k \in \{ 1, \dots, |V(T)| - 1 \}$).
    We show in the following that it then also holds for $i = k + 1$.

    Consider first the case that $v^{(k + 1)}$ is a reflect node.
    By Property~\ref{four} of the induction hypothesis and \Cref{lem:nodeandedge}, there exists an unmarked $2$-leaf in $G^{(k)}$ with label $\ell(v^{(k + 1)})$.
    Now, Property~\ref{one} for $i = k + 1$ follows by Property~\ref{three} of the induction hypothesis (i.e., for $i = k$).
    Property~\ref{two} for $i = k + 1$ also follows from this discussion.

    For proving Property~\ref{three}, let $W$ denote the set of nodes in the maximal connected component containing $v^{(k + 1)}$ in the subgraph of $G^{(k)}$ induced by the unmarked nodes.
    Due to the fact that $G^{(k + 1)}$ is obtained from $G^{(k)}$ by reflection at $v^{(k + 1)}$, we can describe the set of nodes in $G^{(k + 1)}$ as follows.
    For each node in $V(G^{(k)}) \setminus W$ there is one node in $V(G^{(k + 1)})$ with precisely the same label.
    For each node $w \in W \setminus \{v^{(k + 1)} \}$, there are $\Delta$ nodes in $G^{(k + 1)}$ such that $\ell(w)$ is a final substring of the substring of each of the $\Delta$ nodes but there are no two of the $\Delta$ nodes such that the label of the one is a final substring of the label of the other.
    For node $v^{(k + 1)} \in V(G^{k})$, there is precisely one node in $G^{(k + 1)}$ such that $\ell(v^{(k + 1)})$ is a final substring of the label of that node.
    Moreover, the nodes in $V(G^{(k + 1)})$ given in this description are all distinct (but cover all of $V(G^{(k + 1)})$).
    From this description and the fact that Property~\ref{three} is satisfied for $i = k$, we obtain that there are no two nodes in $V(G^{(k + 1)})$ such that the label of the one is a final substring of the label of the other.
    This implies Property~\ref{three} for $i = k + 1$.

    For proving Property~\ref{four}, observe that the reflection at $v^{(k + 1)}$ that produces $G^{(k + 1)}$ from $G^{(k)}$ induces a natural bijection $\beta: \fC^{(k)} \rightarrow \fC^{(k + 1)}$ that simply maps each component $C \in \fC^{(k)}$ to the component in $v^{(k + 1)}$ containing $C$ as a subgraph.
    Moreover, let $\gamma: \fE^{(k)} \rightarrow \fE^{(k + 1)}$ denote the bijection that maps the edge of $T$ with tail $v^{(k + 1)}$ to the edge with head $v^{(k + 1)}$ and any other edge in $\fE^{(k)}$ to itself.
    We claim that $\alpha_{k + 1} := \gamma \circ \alpha_k \circ \beta^{-1}: \fC^{(k + 1)} \rightarrow \fE^{(k + 1)}$ is a bijection as desired in Property~\ref{four} (where $\alpha_k$ is the bijection guaranteed by Property~\ref{four} of the induction hypothesis).
    For proving this claim, due to the definitions of $\beta$ and $\gamma$ (and the induction hypothesis), it suffices to show that, for the component $C \in \fC^{(k)}$ containing $v^{(k + 1)}$,
    \begin{enumerate}
        \item the set of labels of the nodes of $\beta(C)$ that are $2$-leaves of $G^{(k + 1)}$ is precisely $\psi(\gamma(\alpha_k(C)))$, and
        \item the set of labels of the nodes of $\beta(C)$ that are not $2$-leaves of $G^{(k + 1)}$ is precisely $\pi(\gamma(\alpha_k(C)))$.
    \end{enumerate}
    However, the former property follows directly from Property~\ref{five} of the induction hypothesis and the definitions of the reflection operation and $\psi$ in \Cref{def:reflect,def:edgelabeling}, and the latter property follows directly from Property~\ref{five} of the induction hypothesis and the definitions of the reflection operation and $\pi$ in \Cref{def:reflect,def:edgelabelingtwo}.

    Property~\ref{five} follows directly from the definition of the reflection operation in \Cref{def:reflect} and Property~\ref{five} of the induction hypothesis.

    Now consider the case that $v^{(k+1)}$ is a split node.
    By Property~\ref{four} of the induction hypothesis and \Cref{lem:nodeandedgetwo}, there exists an unmarked node in $G^{(k)}$ with label $\ell(v^{(k + 1)})$.
    Now, Property~\ref{one} for $i = k + 1$ follows by Property~\ref{three} of the induction hypothesis (i.e., for $i = k$).
    Property~\ref{two} for $i = k + 1$ also follows from this discussion.
    Moreover, Property~\ref{three} follows directly from the definition of the split operation in \Cref{def:reflect} and Property~\ref{three} of the induction hypothesis.

    For proving Property~\ref{four}, we consider two cases.
    Consider first the case that $v^{(k + 1)}$ is an internal split node, and let $C'$ be the component in $\fC^{(k)}$ containing $v^{(k + 1)}$.
    Let $p$ denote the split index of $v^{(k + 1)}$ in $T$, and let $q$ be the index such that $v^{(q)}$ is the unique ancestor of $v^{(k + 1)}$ in reflect layer $\fR_p$.
    Observe that there is a node $u \in G^{(q)}$ with label $\ell(u) = *\ell(v^{(q)})$, by Property~\ref{two} of the induction hypothesis and the definition of the reflection operation.
    Let $\hat{C}$ denote the component in $\fC^{(q)}$ containing $u$.
    Observe that, by \Cref{lem:compbeforesplit}, Property~\ref{four} of the induction hypothesis, and the definition of the reflection operation, there is a sequence $u = w^{(q)}, w^{(q + 1)}, \dots, w^{(k) = v^{(k + 1)}}$ of nodes in $G^{(q)}, \dots, G^{(k)}$, respectively, such that, for each $q \leq d \leq k - 1$, we have that $w^{(d + 1)}$ is identical\footnote{Recall that we consider $G^{(d + 1)}$ as a supergraph of $G^{(d)}$.} to $w^{(d)}$ or a $g$-copy\footnote{See~\Cref{def:reflect} for the definition of a $g$-copy of a node.} of $w^{(d)}$, for some $1 \leq g \leq \Delta$.
    Moreover, by~\Cref{lem:compbeforesplit}, Property~\ref{four} of the induction hypothesis, and the definitions of $\psi$, $\pi$, and the reflection operation, each node of $C$ can be obtained from some node of $\hat{C}$ by a sequence that uses, for each $d$, precisely the same way to obtain the node in $G^{(d + 1)}$ from the node in $G^{(d)}$ (including the choice of the respective $g$, which are specified by the symbols $q_j$ from \Cref{lem:compbeforesplit}).
    It follows that $C'$ is isomorphic to a subgraph of $\hat{C}$ where the isomorphism preserves the $0$-th to $p$-th symbols of the label of each node (where the preservation comes from the fact that the reflection operation does not change already present symbols of a label when copying a node).
    Hence, by the way in which the reflection operation at $v^{(q)}$ affects the nodes (and in particular their $p$-th symbols) ending up in $G^{(q)}$ and the fact that the split operation at $v^{k + 1}$ turns the image of $u$ under the aforementioned isomorphism into a marked node, we obtain that there are precisely $\Delta$ components $C'_1, \dots, C'_{\Delta} \in \fC^{(k + 1)}$ that are subgraphs of $C'$ and, for each $1 \leq j \leq \Delta$, the label of each node in $C'_j$ has $j$ as its $p$-th symbol.
    Note that the fact that none of these components is empty follows from the fact that $v^{(k + 1)}$ has degree $\Delta$ in $C'$, which in turn follows from the following argumentation:
    
    From \Cref{obs:basicconstree}, \Cref{lem:psiandpileaf}, and Property~\ref{four} of the induction hypothesis, we know that when a $1$-node is marked previous to processing $v^{(k + 1)}$, then the $1$-node did not have any neighbors at that point.
    Hence, no unmarked $2$-leaf of $G^{(k)}$ can contain a $*$ in its label, by the definition of the reflection operation and Property~\ref{five} of the induction hypothesis.
    Since split operations are only performed at nodes with a label that contains a $*$ (Property~\ref{prop:splitstar} of \Cref{def:solidity}), we obtain that $v^{k + 1}$ has degree $\Delta$ in $C'$, by Property~\ref{five} of the induction hypothesis.
    
    Now we are ready to define the desired bijection $\alpha_{k + 1}$.
    Let $e$ denote the edge of $T$ with tail $v^{(k + 1)}$ and, for each $1\leq j \leq \Delta$, let $e_j$ denote the edge whose tail is the $j$-th child of $v^{(k + 1)}$.
    For each component $C \in C^{(k)}$ that does not contain $v^{(k + 1)}$ (which implies that we can consider $C$ also as a component in $C^{(k + 1)}$ as it does not change during the reflection operation at $v^{(k + 1)}$), we set $\alpha_{k + 1}(C) := \alpha_{k}(C)$.
    Moreover, for each $1 \leq j \leq \Delta$, we set $\alpha_{k + 1}(C'_j) = e_j$.
    Now Property~\ref{four} for $i = k + 1$ follows directly from the definitions of $\psi$ and $\pi$, the fact that the label of each node in $C'_j$ has $j$ as its $p$-th symbol (as observed above), and Property~\ref{four} of the induction hypothesis.

    Next, consider the case that $v^{(k + 1)}$ is a leaf split node.
    Then, by \Cref{lem:psiandpileaf}, and Property~\ref{four} of the induction hypothesis, we know that the component in $\fC^{(k)}$ containing $v^{(k + 1)}$ contains no other node than $v^{(k + 1)}$.
    Hence, setting $\alpha_{k + 1}(C) := \alpha_{k}(C)$ for each component $C \in \fC^{(k)}$ that does not contain $v^{(k + 1)}$ yields Property~\ref{four} for $i = k + 1$ (by Property~\ref{four} of the induction hypothesis).

    Finally, we prove Property~\ref{five}.
    The fact that all neighbors of each $2$-node in $G^{(k + 1)}$ are $1$-nodes and vice versa follows directly from Property~\ref{five} of the induction hypothesis and the definition of the split operation.
    Now let $u$ be some $2$-node in some component $C' \in \fC^{(k + 1)}$, and let $C$ be the component in $\fC^{(k)}$ containing $u$.
    By Property~\ref{five} of the induction hypothesis, we know that $u$ is a $2$-leaf of $G^{(k)}$ or has degree $\Delta$ in $C$.
    If $u$ is a $2$-leaf of $G^{(k)}$, then it is also a $2$-leaf of $G^{(k + 1)}$; hence, assume that $u$ has degree $\Delta$ in $C$.
    Again by Property~\ref{five} of the induction hypothesis, we know that all neighbors of $u$ in $G^{(k)}$ are $1$-nodes.
    Thus, by \Cref{obs:basicconstree}, \Cref{lem:psiandpileaf}, and Property~\ref{four} of the induction hypothesis, we know that $v_{(k + 1)}$ is not a neighbor of $u$ in $G^{(k)}$.
    It follows that $u$ has degree $\Delta$ in $C'$, concluding the proof of Property~\ref{five}.

    This concludes the proof that $G_T$ is well-defined for each \emph{fixed} choice of $\vo, \vt, \dots, v^{(|V(T)|)}$.
    Next, we show that $G_T$ is independent of the choice.

    Let $v^{(1)}, v^{(2)}, \dots, v^{(|V(T)|)}$ be an arbitrary sequence as described in \Cref{def:geetee}, and let $1 \leq j \leq |V(T)|$ be an integer such that $v^{(j)}$ is a split node in this sequence.
    For simplicity, let $w$ denote the node in $G^{(j - 1)}$ with label $\ell(v^{(j)})$.
    Observe that, by the proof of Property~\ref{four} of the statement we just proved via induction (and in particular the inductive choice of the bijections $\alpha_i$), the following holds: for any two descendants $u_1, u_2$ of $v^{(j)}$, the two (reflection or split) operations corresponding to $u_1$ and $u_2$ will happen in two parts of the respectively obtained graph that can be reached from $w$ (which is a node that exists in any graph $G^{(p)}$ with $p \geq j - 1$) via two different edges.
    In other words, due to the definition of the reflection and split operations (which only affect components of unmarked nodes, whereas $w$ is marked in each graph $G^{(p)}$ with $p \geq j$), it is irrelevant in which order the two aforementioned operations are executed (as none of the operations has any influence on the effects of the other).
    As this is true for every sequence as described above and every choice of $j$, we obtain that any choice of $v^{(1)}, v^{(2)}, \dots, v^{(|V(T)|)}$ will result in the same graph $G_T$.
\end{proof}

We conclude~\Cref{sec:inputtree} by collecting some useful properties of the sequence $G^{(0)}, \dots, G^{(|V(T)|)}$.
First, by~\Cref{lem:nodeandedge,lem:nodeandedgetwo}, the following observation is implied by the proof of~\Cref{lem:welldef}.

\begin{observation}\label{obs:thisisyourcomponent}
    Let $v^{(1)}, \dots, v^{(|V(T)|)}$ and $G^{(0)}, \dots, G^{(|V(T)|)}$ be as defined in \Cref{def:geetee}.
    Let $2 \leq i \leq |V(T)|$, and let $e$ be the edge with tail $v^{(i)}$.
    Let $C$ denote the maximal connected component in the subgraph of $G^{(i - 1)}$ induced by the unmarked nodes that contains $v^{(i)}$.
    Then,
    \begin{enumerate}
        \item the set of labels of the nodes of $C$ that are $2$-leaves of $G^{(i - 1)}$ is precisely $\psi(e)$, and
        \item the set of labels of the nodes of $C$ that are not $2$-leaves of $G^{(i - 1)}$ is precisely $\pi(e)$.
    \end{enumerate}
\end{observation}

Moreover, there is a simple characterization of the property that two nodes are neighbors in one of the trees $G^{(j)}$ used in the definition of $G_T$, as given in the following lemma.

\begin{lemma}\label{lem:whenneighbors}
    Let $v^{(1)}, \dots, v^{(|V(T)|)}$ and $G^{(0)}, \dots, G^{(|V(T)|)}$ be as defined in \Cref{def:geetee}.
    Let $0 \leq i \leq |V(T)|$ be some integer.
    Let $v \neq w$ be two unmarked nodes of $G^{(i)}$ (which in particular implies that $\len(\ell(v)) = \len(\ell(w))$).
    Then $v$ and $w$ are neighbors in $G^{(i)}$ if and only if, for each $1 \leq j \leq \len(v) - 1$ satisfying $\ell(v), \ell(w) \in \{ 1, \dots, \Delta \}$, we have $\ell(v)_j = \ell(w)_j$. 
\end{lemma}
\begin{proof}
    We prove the lemma by induction on $i$.
    For the base case $i = 0$, the lemma statement follows directly from the definition of $G^{(0)}$.
    Now assume that the lemma statement holds for $i = k$ (where $0 \leq k \leq |V(T)| - 1$).
    We show in the following that it then also holds for $i = k + 1$.

    If $v^{(k + 1)}$ is a split node, then this follows directly from the definition of the split operation (see~\Cref{def:split}).
    Hence, assume that $v^{(k + 1)}$ is a reflect node, and consider two nodes $v \neq w$ of $G^{(k + 1)}$.
    By the definition of the reflection operation (see~\Cref{def:reflect}), there are two (uniquely defined) unmarked nodes $v' \neq w'$ of $G^{(k)}$ such that $\ell(v')_j = \ell(v)_j$ and $\ell(w')_j = \ell(w)_j$, for each $1 \leq j \leq \len(v) - 2$.

    From the definition of the reflection operation, we obtain that $v$ and $w$ are neighbors in $G^{(k + 1)}$ if and only if
    \begin{enumerate}
        \item $v'$ and $w'$ are neighbors in $G^{(k)}$ and
        \item $\ell(v')_{\len(v) - 1} = \ell(w')_{\len(v) - 1}$ or $* \in \{ \ell(v')_{\len(v) - 1}, \ell(w')_{\len(v) - 1} \}$
    \end{enumerate}
    Now the lemma statement for $i = k + 1$ follows from the induction hypothesis.    
\end{proof}

\subsection{A Relation between the Input Trees Induced by \texorpdfstring{$T$}{T} and \texorpdfstring{$F(T)$}{F(T)}}\label{sec:relationtft}
In this section, we will prove a relation between the two input trees $G_T$ and $G_{F(T)}$ that are induced by $T$ and $F(T)$, respectively.
This relation is crucial for the proof of our overall lower bound construction.
Before stating and proving this relation, we introduce some necessary definition.

\begin{definition}[distance-$D$-correct]\label{def:distancedee}
    Let $T$ be a construction tree and $D$ a positive integer.
    Let $\vo, \vt, \dots, v^{(|V(T)|)}$ and $\Gz, \Go, \dots, G^{(|V(T)|)}$ be as in the definition of $G_T$ in \Cref{def:geetee}.
    We say that $T$ is \emph{distance-$D$-correct} if, for each $1 \leq i \leq |V(T)|$ such that $v^{(i)}$ is a split node, each $2$-leaf in $G^{(i - 1)}$ that is in the same component of unmarked nodes as $v^{(i)}$ has distance at least $D$ from $v^{(i)}$ in $G^{(i - 1)}$.
\end{definition}

We remark that analogously to the argumentation in the proof of \Cref{lem:welldef} why $G_T$ is independent of the choice of the nodes $\vo, \vt, \dots, v^{(|V(T)|)}$ we also obtain that whether a construction tree is distance-$D$-correct is independent of the choice of the $\vo, \vt, \dots, v^{(|V(T)|)}$.

Now we are ready to prove the main result of \Cref{sec:relationtft}.

\begin{lemma}\label{lem:fromdto2d}
    Let $D$ be a positive integer and $T$ a distance-$D$-correct $\Delta$-ary construction tree.
    Then $F(T)$ is a distance-$(2D)$-correct $\Delta$-ary construction tree.
\end{lemma}
\begin{proof}
    By~\Cref{lem:iftthenft}, $F(T)$ is a $\Delta$-ary construction tree.
    In the following, we show that it is also distance-$(2D)$-correct.
    
    Let $\vo, \vt, \dots, v^{(|V(F(T))|)}$ and $\Gz, \Go, \dots, G^{(|V(F(T))|)}$ be as in~\Cref{def:geetee}, just for $F(T)$ instead of $T$.
    Recall the definition of the $j$-th child of a node of a construction tree from \Cref{def:solidity}.
    Let $P$ be the root-leaf path in $F(T)$ that, starting from the root $r$, always takes the first child when arriving at a split node.
    We first show ``distance-$(2D)$-correctness for any node on $P$''.

    Let $v^{(i)}$ be an arbitrary split node on $P$.
    By the argumentation in the proof of~\Cref{lem:welldef} guaranteeing independence of the choice of the nodes $\vo, \vt, \dots, v^{(|V(F(T))|)}$, the maximal connected component of unmarked nodes in $G^{(i - 1)}$ containing $v^{(i)}$ is independent of the aforementioned choice and in particular only depends on the nodes (and their labels) in $F(T)$ that lie on the path from $r$ to $v^{(i)}$.
    Hence, w.l.o.g.\ assume that $i$ is the number of nodes on the path from $r$ to $v^{(i)}$ and that the nodes on this path are $r = v^{(1)}, v^{(2)}, \dots, v^{(i)}$ (ordered by increasing distance from $r$).

    Recall that $V_R$ denotes the set of reflect nodes of $T$.
    Let $Q = (w_1, \dots, w_{2|V_R|})$ be the (chronologically ordered) sequence (with repeating elements) of reflect nodes of $T$ visited by the depth-first search traversal used in the definition of $F(T)$ and recall the definition of a late node (see \Cref{def:fcdot}).
    Observe that $v^{(i)} \in L^F_i$.
    Since $v^{(i)}$ is a split node, we know that $w_i$ is a late node.
    Let $v'$ denote the reflect node of $T$ such that $w_i$ is the second occurrence of $v'$ in $Q$, let $r'$ denote the root of $T$, and let $P'$ denote the path from $r'$ to $v'$.
    Moreover, let $r' = v'_1, v'_2, \dots, v'_{|V_R|}$ denote the reflect nodes of $T$ in the order they are visited \emph{for the first time} by the aforementioned depth-first traversal.  

    Recall the notation for the layers of $F(T)$ from~\Cref{def:fcdot}.
    Let $\fK \subseteq \{ 1, \dots, |V_R| \}$ be the set of indices $k$ such that $\fS^F_k$ has layer index $< i$, and $\fK' \subseteq \{ 1, \dots, |V_R| \}$ the set of indices $k'$ such that $\fR^F_{k'}$ has layer index $< i$.
    By the fact that $F(T)$ is well-nested, we know that $\fK \subseteq \fK'$.
    Moreover, by the definition of $F(T)$, a reflect node $v'_j$ lies on $P'$ if and only if $j \in \fK' \setminus \fK$.

    Now consider $G^{(i - 1)}$.
    Let $C$ denote the maximal connected component in the subgraph of $G^{(i - 1)}$ induced by the unmarked nodes that contains $v^{(i)}$.
    Let $u$ be an arbitrary $2$-leaf of $G^{(i - 1)}$ that is contained in $C$.
    Note that if $v^{(i)}$ is a leaf node, then~\Cref{lem:psiandpileaf,obs:thisisyourcomponent} guarantee that no such node $u$ exists and there is nothing to prove.
    Hence, assume in the following that $v^{(i)}$ is an internal split node.
    
    Let $e$ be the edge of $F(T)$ with tail $v^{(i)}$ (and note that $i \neq 1$ by~\Cref{obs:basicconstree}).
    By~\Cref{obs:thisisyourcomponent}, we have $\ell(u) \in \psi(e)$, which, by~\Cref{lem:nodeandedgetwo,obs:thisisyourcomponent}, implies $\ell(u) \neq \ell(v^{(i)})$ and $u \neq v^{(i)}$.
    By the definitions of $\psi$ and $v^{(i)}$, this implies that $\ell(u)_k = 1$ for each $k \in \fK$.
    Moreover, by the definitions of $\psi$ and $F(T)$, we know that $\len(\ell(u)) = \len(\ell(v^{(i)})) = |\fK'| + 1$.

    Observe that the definition of $\psi$ in~\Cref{def:edgelabeling}, together with~\Cref{lem:psiandpileaf}, guarantees that whenever, for some edge $e'$ of a construction tree, a string $z$ is contained in $\psi(e')$, then there is some reflect node $w$ in the subtree of the construction tree hanging from $e'$ such that $z$ is a final substring of $\ell(w)$.
    Applying this observation to construction tree $F(T)$ (with $e' := e$ and $z := \ell(u)$), we obtain that there is some reflect node $w$ in the subtree of $F(T)$ hanging from $v^{(i)}$ (including $v^{(i)}$) such that $\ell(u)$ is a final substring of $\ell(w)$.
    Let $1 \leq d \leq 2|V_R|$ be the index such that $w \in L^F_d$.
    Since $v^{(i)}$ is not a reflect node, we obtain that $d > i$.
    Recall the definition of $Q$, and let $w'$ be the node of $T$ such that $w_d$ is the first occurrence of $w'$ in $Q$.
    Let $x$ denote the split node of $T$ that is an ancestor of both $v'$ and $w'$ and has the greatest distance from the root $r'$ amongst all nodes with this property.
    Observe that, by the definition of $F(T)$, node $w'$ is neither an ancestor nor a descendant of $v'$, which implies that $w' \neq x \neq v'$.
    
    Let $e_x$ denote the edge of $T$ with tail $x$.
    From the definition of $\psi$ (see~\Cref{def:edgelabeling}), it follows that there are strings $y_v, y_w \in \psi(e_x)$ such that $y_v$ is a final substring of $\ell(v')$ and $y_w$ a final substring of $\ell(w')$.    
    Let $j$ be the split index of $x$ and $j'$ the layer index of $x$ in $T$.
    Let $u^{(1)}, \dots, u^{(|V(T)|)}$ be an ordering of the nodes in $V(T)$ as defined in~\Cref{def:geetee} but with the additional property that all nodes of $P'$ appear in the ordering before the first node not contained in $P'$.
    Let $G_T^{(0)}, G_T^{(1)}, \dots, G_T^{(|V(T)|)}$ be the graphs obtained from the aforementioned node sequence in the way described in~\Cref{def:geetee}.
    
    Consider $G_T^{(j' - 1)}$.
    Let $\hat{v}$ be the $2$-leaf of $G_T^{(j' - 1)}$ with label $\ell(\hat{v}) = y_v$ and $\hat{w}$ the $2$-leaf of $G_T^{(j' - 1)}$ with label $\ell(\hat{w}) = y_w$.
    By~\Cref{obs:thisisyourcomponent}, such $2$-leaves $\hat{v}$ and $\hat{w}$ indeed exist in $G_T^{(j' - 1)}$ and are contained in the same maximal connected component of unmarked nodes of $G_T^{(j' - 1)}$ as $x$.
    Moreover, by the fact that $T$ is distance-$D$-correct, we obtain that the distance between $x$ and $\hat{v}$ in $G_T^{(j' - 1)}$ is at least $D$, and that the distance between $x$ and $\hat{w}$ in $G_T^{(j' - 1)}$ is at least $D$.
    

    Observe that the fact that $j$ is the split index of $x$, together with Property~\ref{prop:splitkids} of~\Cref{def:solidity} and the fact that $\ell(\hat{v})$ and $\ell(\hat{w})$ are final substrings of $\ell(v')$ and $\ell(w')$ (respectively), implies that $\ell(\hat{v})_j \neq \ell(\hat{w})_j$.
    By the definitions of $\psi$ and $\pi$ and~\Cref{obs:thisisyourcomponent} (applied to $G^{(j')}$), it follows that the neighbor of $x$ on the path between $x$ and $\hat{v}$ is not the same node as the neighbor of $x$ on the path between $x$ and $\hat{w}$.
    Hence, the distance between $\hat{v}$ and $\hat{w}$ in $G^{(j' - 1)}$ is at least $2D$.
    Let $\hat{v} = x_0, \dots, x_{\alpha} = \hat{w}$ denote the nodes on the path from $\hat{v}$ to $\hat{w}$ in $G^{(j' - 1)}$ (ordered by increasing distance from $\hat{v}$).
    From the discussion above, we know that $\alpha \geq 2D$.

    Recall the depth-first search traversal used in the definition of $F(T)$ (see~\Cref{def:fcdot}), the definition of the nodes $v'_1, \dots, v'_{|V_R|}$, and the definition of $P$.
    Let $\fL' \subseteq \{ 1, \dots, |V_R| \}$ denote the set of all indices $\lambda$ such that the aforementioned depth-first search traversal visits (the reflect node) $v'_{\lambda}$ before (the split node) $x$.
    Let $\fL \subseteq \fL'$ denote the set of all indices $\lambda$ such that $v'_{\lambda}$ does not lie on the path from $r'$ to $x$ in $T$.
    Moreover, if $v'_{|\fL'|}$ lies on the path from $r'$ to $x$ in $T$, then let $v^*$ denote the reflect node on $P$ (in $F(T)$) with reflect index $|\fL'|$; if $v'_{|\fL'|}$ does not lie on the path from $r'$ to $x$ in $T$, then let $v^*$ denote the split node on $P$ (in $F(T)$) with split index $|\fL'|$.
    If $v^*$ is a reflect node, then let $e^*$ be the unique edge with head $v^*$; if $v^*$ is a split node, then let $e^*$ be the edge between $v^*$ and the first child of $v^*$.
    (Note that $e^*$ exists in either case, due to the fact that there are reflect nodes that are descendants of $x$ in $T$, e.g., $v'$ and $w'$.)

    Consider some arbitrary string $y \in \psi(e_x) \cup \pi(e_x)$.
    Let $\gamma(y)$ be the string obtained by taking the $\fL$-padded version of $y$ and then replacing the symbol $\square$ with the symbol $1$ (wherever the symbol $\square$ occurs in the $\fL$-padded version of $y$).
    Then, from~\Cref{obs:rephrasedigestibly} and the definitions of $\psi$, $\pi$, and $F(T)$, we obtain that $\gamma(y) \in \psi(e^*) \cup \pi(e^*)$.
    Consider $x_{\beta}$, for some arbitrary index $0 \leq \beta \leq \alpha$.
    By~\Cref{obs:thisisyourcomponent}, we have $\ell(x_{\beta}) \in \psi(e_x) \cup \pi(e_x)$.
    Hence, from the above discussion, we obtain that $\gamma(\ell(x_{\beta})) \in \psi(e^*) \cup \pi(e^*)$.
    
    Observe that, by the definitions of $F(T)$, $\fL$, and $\fL'$, node $v^*$ lies in layer $L^F_{|\fL'| + |\fL|}$ of $F(T)$.
    As, by definition, $v^*$ lies on $P$, this implies that $v^* = v^{(|\fL'| + |\fL|)}$.    
    Consider $G^{(|\fL'| + |\fL|)}$.
    From the above discussion, the definition of the strings $\gamma(\ell(x_{\beta}))$, and~\Cref{obs:thisisyourcomponent}, it follows that there is a maximal connected component $C'$ of unmarked nodes of $G^{(|\fL'| + |\fL|)}$ such that, for each $0 \leq \beta \leq \alpha$, there is a node of $C'$ with label $\gamma(\ell(x_{\beta}))$, and these nodes are pairwise distinct.
    For each $0 \leq \beta \leq \alpha$, let $\gamma(x_{\beta})$ denote the node with label $\gamma(\ell(x_{\beta}))$.
    By~\Cref{lem:whenneighbors} and the definition of the strings $\gamma(\ell(x_{\beta}))$, we know that, for each $0 \leq \beta \leq \alpha - 1$, nodes $\gamma(x_{\beta})$ and $\gamma(x_{\beta + 1})$ are neighbors in $C'$.
    It follows that $\gamma(x_0)$ and $\gamma(x_{\alpha})$ have distance $\alpha$ in $C'$, and therefore also in $G^{(|\fL'| + |\fL|)}$.
    
    Recall that $\ell(\hat{v})$ is a final substring of $\ell(v')$ and $\ell(\hat{w})$ a final substring of $\ell(w')$.
    By~\Cref{obs:rephrasedigestibly}, the definitions of $v'$, $w'$, $\gamma(x_0)$, and $\gamma(x_{\alpha})$, and the facts that $x_0 = \hat{v}$, $x_{\alpha} = \hat{w}$, and $\fL' \setminus \fL \subseteq \fK' \setminus \fK$, it follows that, for each $\lambda \in (\fL' \setminus \fL) \cup \{ 0 \}$, we have $\ell(\gamma(x_0))_{\lambda} = \ell(v^{(i)})_{\lambda}$ and $\ell(\gamma(x_{\alpha}))_{\lambda} = \ell(w)_{\lambda}$.
    Since, $\fK \subseteq \fL$ and, for each $k \in \fK$, we have $\ell(v^{(i)})_{k} = \ell(w)_{k} = 1$, we also obtain that, for each $\lambda \in \fL$, we have $\ell(\gamma(x_0))_{\lambda} = \ell(v^{(i)})_{\lambda}$ and $\ell(\gamma(x_{\alpha}))_{\lambda} = \ell(w)_{\lambda}$.
    As $\len(\ell(\gamma(x_0))) = \len(\ell(\gamma(x_{\alpha}))) = |\fL'| + 1$, it follows that $\ell(\gamma(x_0))$ is a final substring of $\ell(v^{(i)})$ and $\ell(\gamma(x_{\alpha}))$ a final substring of $\ell(w)$.
    Since $\fL' \subseteq \fK'$ and, for each $k \in \fK'$, we have $\ell(u)_k = \ell(w)_k$, we also obtain that $\ell(\gamma(x_{\alpha}))$ is a final substring of $\ell(u)$.

    Next, observe that, since $v'$ is a descendant of $x$ in $T$ and $v' \neq x$, we have that $v^{(i)}$ is a descendant of $v^* = v^{(|\fL'| + |\fL|)}$ in $F(T)$ and $v^{(i)} \neq v^{(|\fL'| + |\fL|)}$.
    In particular, we obtain that $i >|\fL'| + |\fL|$.    
    Consider $G^{(i)}$.
    The above discussion implies that $G^{(i)}$ is obtained from $G^{(|\fL'| + |\fL|)}$ via the reflection and split operations indicated by the nodes in $F(T)$ on the subpath of $P$ from $v^{(|\fL'| + |\fL| + 1)}$ to $v^{(i)}$.
    

    Let $|\fL'| + |\fL| \leq \eta \leq i - 1$ be some integer, and let $\hat{x}$ and $\overline{x}$ be two nodes of $G^{(\eta + 1)}$.
    Let $\hat{x}'$ and $\overline{x}'$ be the two (uniquely\footnote{That $\hat{x}'$ and $\overline{x}'$ are uniquely defined follows from the definitions of the reflection and split operations and the fact that the set of labels of the nodes of $G^{(\eta)}$ is independent, as shown in the proof of~\Cref{lem:welldef}.} defined) nodes of $G^{(\eta + 1)}$ satisfying that $\ell(\hat{x}')$ is a final substring of $\ell(\hat{x})$ and $\ell(\overline{x}')$ a final substring of $\ell(\overline{x})$.
    Then, the definitions of the reflection and split operations ensure that the distance between $\hat{x}$ and $\overline{x}$ in $G^{(\eta + 1)}$ is at least as large as the distance between $\hat{x}'$ and $\overline{x}'$ in $G^{(\eta)}$.
    Applying this argumentation inductively (and using that $\ell(\gamma(x_0))$ is a final substring of $\ell(v^{(i)})$ and $\ell(\gamma(x_{\alpha}))$ a final substring of $\ell(u)$), we obtain that the distance between $v^{(i)}$ and $u$ in $G^{(i - 1)}$ is at least the distance between $\gamma(x_0)$ and $\gamma(x_{\alpha})$ in $G^{(|\fL'| + |\fL|)}$.
    As the distance between $\gamma(x_0)$ and $\gamma(x_{\alpha})$ in $G^{(|\fL'| + |\fL|)}$ is at least $\alpha$ (as shown above), and $\alpha > 2D$, it follows that the distance between $v^{(i)}$ and $u$ in $G^{(i - 1)}$ is at least $2D$. 
    
    Hence, we have shown ``distance-$(2D)$-correctness for any node on $P$'' in the sense that we proved that for any arbitrary split node $v^{(i)}$ on $P$, each $2$-leaf in $G^{(i - 1)}$ that is in the same component of unmarked nodes as $v^{(i)}$ has distance at least $2D$ from $v^{(i)}$ in $G^{(i - 1)}$.
    Observe that, for each connected component of unmarked nodes, the reflection and split operations (and analogously the labelings produced by $\psi$ and $\pi$) are symmetric regarding the symbols from $\{ 1, \dots, \Delta \}$ (in each position of a string except the one indicated by index $0$).
    Moreover, as observed in the proof of~\Cref{lem:welldef}, different connected components of unmarked nodes
    are independent of each other in the sense that each reflect or split operation affects only the connected component in which it is applied.
    Hence, by symmetry, the aforementioned distance-$(2D)$-correctness result for the nodes on $P$ extends also to the nodes that do not lie on $P$ (and to arbitrary sequences $v^{(1)}, v^{(2)}, \dots, v^{(|V(F(T))|)}$ as defined in~\Cref{def:geetee}).
    In other words, for each $\vo, \vt, \dots, v^{(|V(F(T))|)}$ and $\Gz, \Go, \dots, G^{(|V(F(T))|)}$ (as defined in~\Cref{def:geetee}) and each $1 \leq i \leq |V(F(T))|$ such that $v^{(i)}$ is a split node of $F(T)$, each $2$-leaf in $G^{(i - 1)}$ that is in the same component of unmarked nodes as $v^{(i)}$ has distance at least $2D$ from $v^{(i)}$ in $G^{(i - 1)}$.
    Thus, $F(T)$ is distance-$(2D)$-correct.
\end{proof}

Recall the definition of the sequence $T_2, T_3, \dots$ of construction trees in~\Cref{def:constreesequence}.
From \Cref{lem:fromdto2d} (and the fact that, as is straightforward to verify, $T_2$ is distance-$2$-correct), we obtain the following corollary.

\begin{corollary}\label{cor:distanceofsequence}
	For each positive integer $i$, tree $T_i$ is a distance-$2^{i - 1}$-correct $\Delta$-ary construction tree.
\end{corollary}

\section{The Lower Bound}\label{sec:lowerbound}
While~\Cref{sec:ingredients} took care of collecting the main conceptual and technical ingredients for our overall lower bound construction, \Cref{sec:lowerbound} is devoted to conducting the actual lower bound proof.
We start by defining the notion of a \emph{canonical sequence}.
Similarly to how the actual input tree that we will present to a given online-LOCAL algorithm is obtained by minor adaptations from the tree $G_T$ defined in~\Cref{sec:inputtree}, the input query sequence that we will use for that input tree will be obtained by minor adaptations from the canonical sequence.

\begin{definition}[canonical sequence]\label{def:canonical}
    Let $T$ be a $\Delta$-ary construction tree and $G_T$ the input tree induced by $T$.
    We call a node $v$ of $G_T$ a \emph{mirror node} if during the process that derives $G_T$ from $T$ (see~\Cref{def:geetee}), a reflection operation is performed at $v$. 
    
    Consider the uniquely defined sequence $U = (u_1, \dots, u_{|V_R|})$ of reflect nodes of $T$ with the following properties.
    \begin{enumerate}
        \item Each reflect node of $T$ appears exactly once in the sequence.
        \item For any two nodes $u_i$ and $u_j$ in $U$ satisfying $i \neq j$ it holds that
        \begin{enumerate}
            \item if $u_i$ has smaller layer index in $T$ than $u_j$, then $i < j$, and
            \item if $u_i$ and $u_j$ are in the same reflect layer of $T$ and $\ell(u_i)$ is lexicographically smaller\footnote{That we order nodes in the same layer of $T$ lexicographically in $U$ is an arbitrary choice; any fixed ordering of the nodes of a layer of $T$ works for our purposes.} than $\ell(u_j)$, then $i < j$.
        \end{enumerate}
    \end{enumerate}
    Let $W = (w_1, \dots, w_{|V_R|})$ be the sequence of mirror nodes of $G_T$ satisfying that, for each $1 \leq i \leq |V_R|$, label $\ell(u_i)$ is a final substring of $\ell(w_i)$.
    We call $W$ the \emph{canoncial sequence for $G_T$}.
\end{definition}

Note that the definition of $G_T$ in~\Cref{def:geetee} (in particular the definitions of the reflection and split operations), together with Property~\ref{prop:indepandclearing} of~\Cref{def:solidity}, ensures that $W$ is uniquely defined (and exists).

Next, we prove a lemma that is crucial for defining the aforementioned ``adaptations'' that we will use to derive the actual input tree for our lower bound construction from $G_T$.
This lemma ascertains that certain subgraphs of $G_T$ are isomorphic.

\begin{lemma}\label{lem:symmetricview}
    Let $T$, $G_T$, $U$, and $W$ be defined as in~\Cref{def:canonical}.
    Consider some arbitrary node $w_i \in W$ with $i \neq |V_R|$ and let $D$ be some positive integer such that $T$ is distance-$D$-correct.
    Let $W_i$ be the set of all nodes $w$ of $G_T$ such that
    \begin{enumerate}
        \item on the path from $w_i$ to $w$ (including $w_i$ and $w$) in $G_T$, there is no node that is contained in $\{ w_1, \dots, w_{i - 1} \}$, and
        \item\label{item:onlinewithoutonline} for some positive integer $k$, there exists a sequence $(w_i, w_{j_1}, w_{j_2}, \dots, w_{j_k}, w)$ of (not necessarily distinct) nodes such that
        \begin{enumerate}
            \item for each $1 \leq p \leq k$, we have $1 \leq j_p \leq i$,
            \item the distance between $w_i$ and $w_{j_1}$ in $G_T$ is at most $2D - 2$,
            \item for each $1 \leq p \leq k - 1$, the distance between $w_{j_p}$ and $w_{j_{p + 1}}$ in $G_T$ is at most $2D - 2$, and
            \item the distance between $w_{j_k}$ and $w$ in $G_T$ is at most $D - 1$.
        \end{enumerate}
    \end{enumerate}
    Let $G[W_i]$ be the subgraph of $G_T$ induced by $W_i$ (which, by definition, is connected).
    Moreover, for any edge $e$ incident to $w_i$, let $G_e[W_i]$ be the subgraph of $G[W_i]$ induced by the set of nodes consisting of $w_i$ and all nodes of $W_i$ reachable from $w_i$ via a path whose first edge is $e$.
    Then, for any two edges $e, e'$ incident to $w_i$, the graphs $G_e[W_i]$ and $G_{e'}[W_i]$ are isomorphic and there exists an isomorphism between these two graphs that maps $w_i$ to itself and preserves port numbers except possibly at $w_i$. 
\end{lemma}
\begin{proof}
    Let $v^{(1)}, v^{(2)}, \dots, v^{(|V(T)|)}$ and $G^{(0)}, G^{(1)}, \dots, G^{(|V(T)|)}$ be as defined in~\Cref{def:geetee} with the additional property\footnote{The existence of such node and graph sequences with this additional property directly follows from the definition of $U$.} that, for any two reflect nodes $v^{(q)}, v^{(q')}$ with $q < q'$, node $v^{(q')}$ appears before $v^{(q)}$ in $U$.
    Let $i'$ be the index such that $v^{(i')} = u_i$.
    Consider $G^{(i')}$.
    Let $W'_{i'}$ denote the set of nodes of $G^{(i')}$ that lie in the same maximal connected component of unmarked nodes as $v^{(i')}$.
    Due to the fact that $G_T = G^{(|V(T)|)}$ is a supergraph\footnote{Recall that we consider the graph obtained by a reflection or split operation as a supergraph of the graph before the operation, see~\Cref{def:reflect,def:split}.} of $G^{(i')}$, we can consider $W'_{i'}$ to be a subset of the nodes of $G_T$.
    
    Let $G[W'_{i'}]$ be the subgraph of $G_T$ induced by $W'_{i'}$.
    Moreover, for any edge $e$ incident to $v^{(i')}$, let $G_e[W'_{i'}]$ be the subgraph of $G[W'_{i'}]$ induced by the set of nodes consisting of $v^{(i')}$ and all nodes of $G[W'_{i'}]$ reachable from $v^{(i')}$ via a path whose first edge is $e$.
    Now, it directly follows from the definition of the reflection operation and the fact that $G^{(i')}$ is the graph obtained from $G^{(i' - 1)}$ by reflection at $v^{(i')}$ that, for any two edges $e, e'$ incident to $v^{(i')}$, the graphs $G_e[W'_{i'}]$ and $G_{e'}[W'_{i'}]$ are isomorphic and there exists an isomorphism between these two graphs that maps $v^{(i')}$ to itself and preserves port numbers except possibly at $v^{(i')}$.

    Observe that, by the definition of $G_T$ (in particular by the label manipulation happening during a reflection operation), the fact that $v^{(i')} = u_i$ in $T$ implies that $w_i = v^{(i')}$ in $G_T$.
    Moreover, observe that the construction of $G_T$, together with the definition of the canonical sequence $W$, ensures that, for any node $w' \in V(G_T) \setminus V(G^{(i')})$, the path from $w_i$ to $w'$ contains a node from $\{ w_1, \dots, w_{i - 1} \}$ or a node that is marked in $G^{(i')}$.
    Observe that, due to the fact that $T$ is distance-$D$-correct, the distance in $G_T$ between any node in $\{ w_1, \dots, w_i \}$ and any node that is marked in $G^{(i')}$ is at least $D$.
    Hence, by the definition of $W_i$, we obtain $W_i \subseteq W'_{i'}$.

    Observe that, due to Property~\ref{prop:indepandclearing} of \Cref{def:solidity} and the definitions of $W$ and $G_T$, a node from $W'_{i'}$ is contained in $\{ w_1, \dots, w_{i - 1} \}$ if and only if it is also a $2$-leaf in $G^{(i')}$.
    Now, analogously to above, the fact that $G^{(i')}$ is obtained from $G^{(i' - 1)}$ by reflection at $v^{(i')}$ implies that, for any two edges $e, e'$ incident to $w_i = v^{(i')}$, the graphs $G_e[W_{i}]$ and $G_{e'}[W_{i}]$ are isomorphic and there exists an isomorphism between these two graphs that maps $w_i$ to itself and preserves port numbers except possibly at $w_i$.
\end{proof}

As mentioned before, the lower bound instance that we present to a given algorithm depends on the respective algorithm.
In~\Cref{lem:lbofd}, informally speaking, we show that if an algorithm would solve sinkless orientation w.h.p. with too small complexity, then there is an input instance on which the algorithm fails with a relatively large probability.
This lemma constitutes the main ingredient for our lower bound proof for sinkless orientation presented in~\Cref{thm:solb}.
Before proving~\Cref{lem:lbofd}, we introduce the key notion of the \emph{smallest frequent edge} incident to some node $v$, which is a certain edge that the considered algorithm orients away from $v$ with a large probability, relatively speaking. 

\begin{definition}[smallest frequent edge]\label{def:smallfreq}
    Let $\fA$ be a sinkless orientation algorithm in the randomized online-LOCAL model that, when presented a node, always (i.e., for any random bits it may use) chooses at least one incident edge to be outgoing.\footnote{Note that since randomized online-LOCAL algorithms may fail with a small probability, such an algorithm may in principle decide to choose all incident edges to be incoming with small probability.}
    Consider an arbitrary sequence of nodes $(u_1, u_2, \dots)$ that is presented to $\fA$ and let $i \geq 2$ be some integer such that $u_i$ is contained in this sequence.
    Let $e_1, \dots, e_{\deg(u_i)}$ be the edges incident to $u_i$ corresponding to port numbers $1, \dots, \deg(u_i)$, respectively.
    Consider the point of the execution of $\fA$ when $\fA$ has finished deciding on the outputs of the edges incident to $u_1, \dots, u_{i - 1}$ and is presented with $u_i$.
    Consider any fixed set of possible behaviors $\fA$ might have displayed (i.e., any fixed set of overall outputs it might have produced) and let $\fE$ be the probabilistic event that $\fA$ showed one of these behaviors in the execution so far.
    Assume that $\fE$ occurs with nonzero probability.
    Let $e_q$ be the edge with minimal index $q$ such that $\fA$ chooses $e_q$ to be oriented away from $u_i$ with probability at least $1/\Delta$, conditioned on $\fE$.
    Then, we call $e_q$ the \emph{smallest frequent edge incident to $u_i$ conditioned on $\fE$}.
    For the special case of $i = 1$, we call the edge $e_q$ with minimal index $q$ such that $\fA$ chooses $e_q$ to be oriented away from $u_i$ with probability at least $1/\Delta$ (without any conditioning) the \emph{smallest frequent edge incident to $u_i$}.
\end{definition}

\begin{lemma}\label{lem:lbofd}
    Let $D$ and $\Delta \geq 3$ be positive integers.
    Let $T$ be a distance-$D$-correct $\Delta$-ary construction tree and let $V_R$ denote the set of reflect nodes of $T$.
    Let $\fA$ be an algorithm for sinkless orientation in the randomized online-LOCAL model with complexity $D - 1$.
    Then for each $n \geq \Delta \cdot |V(G_T)|$ there exists an $n$-node tree $\fT$ (together with a suitable query sequence) which $\fA$ solves incorrectly with probability at least $1/\Delta^{|V_R|}$.
\end{lemma}
\begin{proof}
    We can assume w.l.o.g.\ that whenever a node is being presented to it, Algorithm $\fA$ (which has to output an orientation for each incident edge) will orient at least one edge incident away from the node.\footnote{If $\fA$ does not satisfy this property, simply replace $\fA$ by an algorithm $\fA'$ that behaves like $\fA$ except that when $\fA$ would orient all edges towards a presented node (for the first time), $\fA'$ chooses one incident edge to be outgoing (and after that behaves in an arbitrary manner that guarantees the aforementioned property). It follows that, on any instance, the failure probability of $\fA'$ is at least as small as the failure probability of $\fA$, which implies that it suffices to prove the lemma statement for $\fA'$.}
    Let $n$ be some integer satisfying $n \geq \Delta \cdot |V(G_T)|$.
    We first show how to construct a tree $\fT$ as mentioned in the lemma and then show why $\fA$ solves $\fT$ incorrectly with probability at least $1/\Delta^{|V_R|}$.
    To prove the latter part, we will show how to construct (together with $\fT$) a sequence $\fW$ of nodes of $\fT$ such that when $\fA$ is presented with the nodes of $\fW$ (in the order given by $\fW$) on $\fT$, the decisions made by $\fA$ will lead to an incorrect output with probability at least $1/\Delta^{|V_R|}$.
    
    On a high level, $\fT$ is obtained from $G_T$ by (a minor initial step for technical reasons and) a number of operations that adapt the structure of $G_T$ to the design of Algorithm $\fA$.
    Essentially, we aim at tweaking $G_T$ so that on the obtained tree, for each decision that $\fA$ takes when presented with $\fW$, the choice of $\fA$ is ``reasonably bad'' with large probability (relatively speaking).
    Sequence $\fW$ is obtained by tweaking the canoncial sequence for $G_T$ in a manner aligning with the tweaking of $G_T$.

    Let $W = (w_1, \dots, w_{|V_R|})$ be the canonical sequence for $G_T$.
    We construct $\fT$ from $G_T$ via a sequence of graphs $G_T = G_{-1}, G_0, \dots, G_{|V_R| - 1} = \fT$.
    At the same time, we construct $\fW$ from $W$ via a sequence of nodes sequences $W = W^{(0)}, \dots, W^{(|V_R| - 1)} = \fW$, where, for each $1 \leq g \leq |V_R| - 1$, we denote the elements of $W^{(g)}$ by $W^{(g)}_1, \dots, W^{(g)}_{|V_R|}$ (given in the order in which they appear in $W^{(g)}$).
    $G_0$ is simply the graph obtained from $G_T$ by attaching to each node $v$ of $G_T$ precisely $\Delta - \deg(v)$ (additional) leaf nodes and then attaching a path of suitable length to one of the new leaf nodes to increase the number of nodes to precisely $n$ (which is possible due to the requirement that $n \geq \Delta \cdot |V(G_T)|$).
    Hence, every node of $G_0$ that is not one of the added nodes has degree $\Delta$ in $G_0$.
    The port numbers for the new edges are assigned in an arbitrary correct manner that does not change already fixed port numbers.
    
    For the construction of $G_1, \dots, G_{|V_R| - 1}$, we emphasize that while the nodes of $G_T$ have been assigned labels (and nodes in the graphs $G_0, \dots, G_{|V_R| - 1}$ could be considered to inherit node labels), these labels are only for the purpose of being able to address the nodes (in particular for defining the canonical sequence for $G_T$); the labels are not available to Algorithm $\fA$.
    When asked to provide some output at a node, as following from the description of the randomized online-LOCAL model, Algorithm $\fA$ (besides remembering its previous decisions) has access only to the structure and port numbers of the parts of the input graph that are at distance at most the complexity of $\fA$ from some node that has been already given to $\fA$ (including the node for which $\fA$ has to make the current decision).

    Let $1 \leq i \leq |V_R| - 1$.
    Assume that $G_{-1}, G_0, \dots, G_{i - 1}$ and $W^{(0)}, \dots, W^{(i - 1)}$  have already been defined.
    We define $G_i$ and $W^{(i)}$ as follows.

    Let $e_1, \dots, e_{\Delta}$ be the edges incident to $w^{(i - 1)}_i$ corresponding to port numbers $1, \dots, \Delta$, respectively.
    Consider the execution of $\fA$ on $G^{(i - 1)}$ with sequence $W^{(i - 1)}$.
    Let $\fE_1$ be the probabilistic event that $\fA$, when being presented with $w^{(i - 1)}_1$, chooses the smallest frequent edge incident to $w^{(i - 1)}_1$ to be outgoing.
    For each $2 \leq d \leq i - 1$ let $\fE_d$ be the probabilistic event\footnote{Note that the recursive definition guarantees that we condition only on events that occur with nonzero probability, which implies the applicability of~\Cref{def:smallfreq}.} that 
    \begin{enumerate}
        \item when being presented with $w^{(i - 1)}_1$, Algorithm $\fA$ chooses the smallest frequent edge incident to $w^{(i - 1)}_1$ to be outgoing, and
        \item for each $2 \leq d' \leq d$, Algorithm $\fA$, when presented with $w^{(i - 1)}_{d'}$, chooses the smallest frequent edge incident to $w^{(i - 1)}_{d'}$ conditioned on $\fE_{d' - 1}$ to be outgoing.
    \end{enumerate}
    Let $e_q$ be the smallest frequent edge incident to $w^{(i - 1)}_i$ conditioned on $\fE_{i - 1}$.
    Let $G'$ denote the (connected) subgraph of $G$ consisting of all nodes that
    \begin{enumerate}
        \item have not been presented to $\fA$ so far, and
        \item can be reached from $w^{(i - 1)}_i$ via a path that uses only nodes that have not been presented to $\fA$ so far (excluding $w^{(i - 1)}_i$ itself) and uses only edges that $\fA$ has already seen\footnote{See~\Cref{def:onlinelocalmodel} for when an online-LOCAL algorithm has seen some edge.} (after $w^{(i - 1)}_i$ has been presented to $\fA$).
    \end{enumerate}
    For each $1 \leq p \leq \Delta$, let $G'[p]$ denote the (connected) subgraph of $G'$ induced by those nodes that can be reached from $w^{(i - 1)}_i$ via the edge incident to $w^{(i - 1)}_i$ that has been assigned port number $p$ by $w^{(i - 1)}_i$ (excluding $w^{(i - 1)}_i$ itself).
    Let $\xi$ be an isomorphism from $G'$ to itself such that \begin{enumerate}
        \item every node in $\{ w^{(i - 1)}_i \} \cup \bigcup_{1 \leq p \leq \Delta, 1 \neq p \neq q} V(G'[p])$ is mapped to itself by $\xi$,
        \item every node in $V(G'[1])$ is mapped to some node in $V(G'[q])$ by $\xi$ and vice versa,
        \item $\xi(\xi(u)) = u$ for each node $u \in V(G'[1]) \cup V(G'[q])$, and
        \item $\xi$ preserves port numbers except possibly regarding the two port numbers assigned by $w^{(i - 1)}_i$ to $e_1$ and $e_q$.
    \end{enumerate}
    Now, $G_i$ is defined as the graph obtained from $G_{i - 1}$ by the following operation.
    For each edge $e$ of $G_{i - 1}$ between some node $u \in V(G'[1]) \cup V(G'[q])$ and some node $u' \in V(G) \setminus V(G')$, replace $e = \{ u, u' \}$ by a new edge $e' = \{ \xi(u), u' \}$ (where the port number assigned to $e'$ by $u'$ is identical to the port number assigned to $e$ by $u'$ and the port number assigned to $e'$ by $\xi(u)$ is identical to the port number assigned to $e$ by $u$).
    Similarly, $W^{(i)}$ is obtained from $W^{(i - 1)}$ by the following operations.
    For each node $w^{(i)}_g$ that is contained in $G'$, set $w^{(i)}_g := w^{(i - 1)}_g$ for each $1 \leq g \leq i$ and $w^{(i)}_g := \xi(w^{(i - 1)}_g)$ for each $i + 1 \leq g \leq |V_R|$.
    For each node $w^{(i)}_g$ that is not contained in $G'$, set $w^{(i)}_g := w^{(i - 1)}_g$.
    
    In the following we argue that the defined graphs and node sequences are well-defined.
    In particular, we show that an isomorphism as described above indeed exists.

    Consider for a moment the scenario that for each node that $\fA$ is presented with, $\fA$ decides to orient the edge that is connected to the considered node via port $1$ outwards with probability at least $1/\Delta$.
    In this case, the desired isomorphism trivially exists and we obtain $\fT = G_{|V_R| - 1} = \dots = G_{0}$ and $\fW = W^{(|V_R| - 1)} = \dots = W^{(0)} = W$.
    Moreover, two further properties hold:
    First, the fact that $T$ is distance-$D$-correct, together with the definition of $W$ (and the fact that $\fA$ has complexity $D - 1$), implies that, after being presented all nodes in $\fW = W$ on $\fT = G_0$, Algorithm $\fA$ has seen only nodes that belong to $G_T$, all of which have degree $\Delta$ in $\fT$.
    Second, the definition\footnote{In particular, it is crucial that in the definition of the reflection operation (see~\Cref{def:reflect}), the original graph $H_1$ is the one whose nodes can be reached \emph{via port $1$} from the node at which the reflection takes place.} of $G_T$ implies that, whenever a node $w_i \in \fW$ is presented to $\fA$, all nodes of $\fW$ that are presented later to $\fA$ (i.e., that have index greater than $i$ in $\fW = W$) lie in the part of $\fT$ that can be reached from $w_i$ via the smallest frequent edge incident to $w_i$ (with the usual conditioning), i.e., via the edge at port $1$ in the currently considered case.
    
    Now consider the general case that does not restrict the decisions of $\fA$ to always orient the incident edge at port $1$ outwards.
    From the definitions of the $G_i$ and $W^{(i)}$ and \Cref{lem:symmetricview}, it follows inductively that the aforementioned two properties (where the restricted $\fW = W$, $\fT = G_0$ etc. are replaced by the $\fW$, $\fT$ etc. respectively induced by the considered $\fA$) also hold in the general case, and that the desired isomorphism exists.
    (Note that Property~\ref{item:onlinewithoutonline} of~\Cref{lem:symmetricview} essentially provides a reformulation of ``the view of an online-LOCAL algorithm with complexity $D - 1$'' that does not need to refer to such an algorithm.)

    Next, we prove that when $\fA$ is executed on $\fT$ with\footnote{Technically, in the randomized online-LOCAL model an algorithm is presented with \emph{all} nodes (in some order) but if the algorithm outputs an incorrect partial output (that cannot be completed to a correct global output) with some probability already for a sequence of nodes that only contains a subset of the nodes of the input graph, then, by a simple completion argument, there must also exist a full sequence (containing all nodes) for which the algorithm fails with at least the same probability. As our goal is to prove a lower bound, we can therefore also consider such partial sequences as valid input query sequences.} node sequence $\fW$, the failure probability of $\fA$ is at least $1/\Delta^{|V_R|}$.
    To this end, consider the events $\fE_d$ discussed above, but defined (analogously) for $\fW = W^{(|V_R| - 1)}$ instead of for $W^{(i - 1)}$ (with $1 \leq d \leq |V_R| - 1$).
    By the above discussion, it follows that
    \begin{enumerate}
        \item the probability of $\fE_d$ to occur is at least $1/\Delta^d$, for each $1 \leq d \leq |V_R| - 1$, and
        \item when $\fE_{|V_R| - 1}$ occurs, then, for each $1 \leq d' \leq |V_R| - 1$, the first edge on the path from $w^{(|V_R| - 1)}_{d'}$ to $w^{(|V_R| - 1)}_{|V_R|}$ in $\fT$ is oriented towards $w^{(|V_R| - 1)}_{|V_R|}$.
    \end{enumerate}
    In particular, the probability that, for each $1 \leq d' \leq |V_R| - 1$, the first edge on the path from $w^{(|V_R| - 1)}_{d'}$ to $w^{(|V_R| - 1)}_{|V_R|}$ in $\fT$ is oriented towards $w^{(|V_R| - 1)}_{|V_R|}$ is at least $1/\Delta^{|V_R| - 1}$.

    Now consider the last node $w_{|V_R|}$ in the canonical sequence $W$ for $G_T$.
    Observe that Property~\ref{prop:indepandclearing} of~\Cref{def:solidity} (together with the definitions of $W$ and the reflection operation) implies that all neighbors of $w_{|V_R|}$ in $G_T$ (and therefore also in $G_0$) appear in $W$ (and necessarily before $w_{|V_R|}$).
    Observe further that the defined ``isomorphic operations'' performed to obtain $\fW = W^{|V_R| - 1}$ from $W$ and $\fT$ from $G_0$ preserve this property: for each $1 \leq g \leq |V_R| - 1$, all neighbors of $w^{(g)}_{|V_R|}$ in $G_{g}$ appear in $W^{(g)}$ (before $w^{(g)}_{|V_R|}$).
    In particular, all neighbors of $w^{(|V_R| - 1)}_{|V_R|}$ in $G_{|V_R| - 1} = \fT$ appear in $W^{(|V_R| - 1)}$ (before $w^{(|V_R| - 1)}_{|V_R|}$).

    Hence, the probability that $\fA$ orients all edges incident to $w^{(|V_R| - 1)}_{|V_R|}$ towards $w^{(|V_R| - 1)}_{|V_R|}$ is at least $1/\Delta^{|V_R| - 1}$.
    This implies that $\fA$ solves $\fT$ incorrectly with probability at least $1/\Delta^{|V_R|}$.
    Since the operations by which $\fT$ is obtained from $G_0$ do not change the number of nodes (and $|V(G_0)| = n$), the lemma statement follows.
\end{proof}

Now we have all the ingredients to prove our lower bound for sinkless orientation in the randomized online-LOCAL model (which implies all the results stated in~\Cref{cor:lllandso}, including the same lower bound for the complexity of the distributed LLL in quantum-LOCAL).

\solb*
\begin{proof}
	Set $\Delta := 4$ and recall the compact power tower notation introduced in~\Cref{def:powertower}.
	Let $n > 4^{P_4(2, 5)}$ be some arbitrary integer.
	Let $i$ be the largest integer such that $n > 4^{P_4(i, 5)}$.

	Consider the $4$-ary construction tree $T_i$.
	By~\Cref{cor:distanceofsequence}, $T_i$ is distance-$2^{i - 1}$-correct.
	Moreover, by~\Cref{lem:numbernodesub}, the number of nodes of $T_i$ (and therefore also the number of reflect nodes of $T_i$) is at most $P_4(i, 5)$.

	Observe that~\Cref{obs:thisisyourcomponent}, together with~\Cref{lem:psiandpileaf}, implies that every node of $G_{T_i}$ is marked.
    By the construction of $G_{T_i}$, it follows that for each node $v$ of $G_{T_i}$, there exists a unique node of $T_i$ that corresponds to the operation of marking $v$ (during the process of operations used to construct $G_{T_i}$), and all of these unique nodes are distinct.
    Hence, $|V(G_{T_i})| \leq |V(T_i)|$.
    This in particular also implies that $n > 4^{|V(T_i)|} \geq 4^{|V(G_{T_i})|} \geq 4 \cdot |V(G_{T_i})| = \Delta \cdot |V(G_{T_i})|$. 

    Now let $\fA$ be an arbitrary algorithm that solves sinkless orientation on $n$-node graphs in the randomized online-LOCAL model with probability at least $1 - 1/n$.
    By~\Cref{lem:lbofd} (and the above discussion), the complexity of $\fA$ cannot be smaller than $2^{i - 1}$ as otherwise, on some $n$-node tree $\fT$, Algorithm $\fA$ would produce an incorrect output with probability at least $1/4^{P_4(i, 5)} > 1/n$.

    Observe that the definition of $i$ (together with the fact that $2^{2^x} > 4^x$ for any sufficiently large real number $x$) implies that $i \in \Theta(\log^* n)$.
    We obtain that $\fA$ has complexity in $2^{\Omega(\log^* n)}$, as desired.
    Moreover, by the construction of $\fT$ in the proof of~\Cref{lem:lbofd}, we also obtain that the lower bound holds already on trees with maximum degree $4$ in which no node has degree $3$.
\end{proof}

\section*{Acknowledgements}
We would like to thank Jukka Suomela for helpful discussions regarding the online-LOCAL model and proposing the studied problem. We would also like to thank Francesco d'Amore for notifying us about the fact that our approach for lifting the deterministic to the randomized online-LOCAL lower bound can be replaced by a simple black-box application of~\cite[Lemma 7.6, arxiv version]{akbari2025online}

Funded by the European Union. Views and opinions expressed are however those of the author(s) only and do not necessarily reflect those of the European Union or the European Research Council. Neither the European Union nor the granting authority can be held responsible for them. This work is supported by ERC grant \href{https://doi.org/10.3030/101162747}{OLA-TOPSENS} (grant agreement number 101162747) under the Horizon Europe funding programme.

\urlstyle{same}
\bibliographystyle{plain}
\bibliography{references}

\end{document}